\documentclass[a4paper,11pt]{article}
\usepackage{mathrsfs} \usepackage{amsmath} \usepackage{amssymb} \usepackage{slashed}
\usepackage{graphicx} \usepackage{caption}
\begin{document}
\newcommand{\ba}{\begin{eqnarray}} \newcommand{\ea}{\end{eqnarray}}
\newcommand{\be}{\begin{equation}} \newcommand{\ee}{\end{equation}}
\renewcommand{\figurename}{Figure}
\renewcommand{\thefootnote}{\fnsymbol{footnote}}

\vspace*{1cm}
\begin{center}
 {\Large\textbf{A Unified Model of Particle Physics and Cosmology: Origin of Inflation, Baryogenesis, Neutrino Mass, CDM and Dark Energy}}

 \vspace{1cm}
 \textbf{Wei-Min Yang}

 \vspace{0.3cm}
 \emph{Department of Modern Physics, University of Science and Technology of China}

 \emph{Hefei 230026, People's Republic of China}

 \emph{E-mail: wmyang@ustc.edu.cn}
\end{center}

\vspace{1cm}
\noindent\textbf{Abstract}:
 I propose a unified model of particle physic and cosmology based on both a new extension of the standard particle model and the fundamental principle of the standard cosmology. Beyond the SM, I introduce several species of dark fields with the dark symmetry of $Z_{2}\otimes Z'_{2}$, the inflation field, CDM and DE are of course included among them. The model can coherently and completely describe the origin and evolution of the universe from the primordial inflation to the reheating, to the baryogenesis, to the current CDM and DE, and also the neutrino mass generation mechanism. For the energy evolution of each phase, I give its dynamical system of equations, and solve them by some special techniques. The numerical results can clearly show how each evolution is successfully implemented, furthermore, the model demonstrates that the DE genesis is purely from the CDM condensation, which is essentially a reverse process of the primordial slow-roll inflation, there is not the so-called ``cosmological constant energy". By use of fewer input parameters, the model not only exactly reproduces the measured inflationary data and the current energy density budget, but also finely predicts such important values as the tensor-to-scalar ratio $r_{0.05}\approx1.68\times10^{-6}$, the inflaton mass $M_{\Phi}\approx5.05\times10^{10}$ GeV, the reheating temperature $T_{re}\approx8.29\times10^{10}$ GeV, the CDM mass $M_{S}\approx256$ GeV, $\eta_{B}\approx6.14\times10^{-10}$ and $h\approx0.73$, in particular, the model smoothly clarifies and eliminates the ``Hubble tension". In short, the unified model newly establishes the deeper and closer relations between particle physics and cosmology, therefore we expect the ongoing and future experiments to test the model.

\vspace{1cm}
\noindent\textbf{Keywords}: beyond standard model; inflation; baryogenesis; neutrino mass;

            \hspace{1.45cm} dark matter; dark energy

\newpage
\noindent\textbf{I. Introduction}

\vspace{0.3cm}
 The standard model of particle physics (SM) and the $\Lambda$CDM model of cosmology (namely the cosmological constant energy + the cold dark matter) together have successfully accounted for a great deal of the cosmic observations from the BBN era to the present day \cite{1}, but they can not answer the origin of the hot big bang of the universe \cite{2}, namely what happened before the standard hot expansion, and also can not effectively address the origins of the baryon asymmetry \cite{3},  the cold dark matter (CDM) \cite{4}, and the dark energy \cite{5}, in addition, the generation of the sub-eV neutrino mass is as yet a puzzle \cite{6}. At present the theoretical and experimental investigations have clearly indicated that the very early universe certainly underwent the inflation phase and the followed reheating one \cite{7}, these two processes not only provide the initial conditions of the hot evolution of the universe, but also make the universe matter genesis \cite{8}, nevertheless, their evolution dynamics have been unestablished as yet, their relations with particle physics are also unknown. To solve all of the above problems, we have to seek an underlying theory beyond the SM and $\Lambda$CDM, therefore this becomes the most challenging research for particle physics and cosmology. This aspect is currently attracting more and more attentions of theoretical and experimental physicists \cite{9}.

 In fact, there have been numerous theories about the explanations of the inflation, dark energy, dark matter, baryon asymmetry and neutrino mass, which include many extensions of the SM \cite{10}, some paradigms of the inflation and reheating \cite{11}, some special models of the dark energy \cite{12}, even some models based on the non-standard gravity \cite{13}, many candidates of the CDM \cite{14}, the modified Newtonian dynamics \cite{15}, some mechanisms of baryogenesis and leptogenesis \cite{16}, and many models of neutrino mass generation \cite{17}. However, a wide variety of these proposals have a common shortcoming, namely they are only aiming at one or two specific aspects of the above-mentioned universe phenomena rather than considering internal connections among them, in other words, these phenomena are dealt in isolation without regard to their integration and coherence in the universe evolution, this is obviously unnatural and inadvisable because the uniqueness of the universe origin and evolution destines that there are surely some internal relations among these universe ingredients. However, the vast majority of these models have been ruled out by the recent data and analyses \cite{18}.

 At the present day, by means of the analyses for the power spectra of the anisotropic and polarized temperature of the cosmic microwave background (CMB) \cite{19}, we have obtained the following data of such inflationary quantities as the scalar power spectra, the scalar spectral index, the running of the spectral index, and the tensor-to-scalar ratio. On the other hand, from the global analyses of cosmology which includes CMB, BBN, structure formation, gravitational lenses, particle physic experiments, etc. \cite{20}, we have extracted the following universe data, the CDM density, the dark energy density, the ratio of the baryon number density to the photon one, and the neutrino mass sum. The present optimum values of these cosmological data are given as follows \cite{1},
\begin{alignat}{1}
 &ln(10^{10}\Delta^{2}_{R})\approx3.044\,,\hspace{0.5cm} n_{s}\approx0.965\,,\hspace{0.5cm}
   \frac{dn_{s}}{dlnk}\approx-0.004\,,\hspace{0.5cm} r_{0.05}<0.036\,,\nonumber\\
 &\Omega_{CDM}\approx\frac{0.12}{h^{2}}\,,\hspace{0.5cm} \Omega_{DE}\approx1-\frac{0.143}{h^{2}}\,,\hspace{0.5cm}
   \eta_{B}\approx6.14\times10^{-10},\hspace{0.5cm} \sum_{i}m_{\nu_{i}}\sim0.1\:\mathrm{eV},
\end{alignat}
 where $h$ is the current scaling factor for Hubble expansion rate, namely $H_{0}=100h$ $\mathrm{km}\mathrm{s}^{-1}\mathrm{Mpc}^{-1}$. There are currently two distinct values for $h$, $h\approx0.73$ is directly measured by the distance ladder approach at the low red-shift, while $h\approx0.674$ is derived from the Planck data of the CMB (which are created at the high red-shift) by assuming the $\Lambda$CDM model, this inconsistency is called as the ``Hubble tension" \cite{21a}. Undoubtedly, these data of Eq. (1)  contain the key information of the universe origin and evolution, therefore any one successful theory of particle physics and cosmology has to confront them unavoidably, so Eq. (1) severely constrains new model builds \cite{21b}.

 Based on the universe concordance and the nature unification, I attempt to build a unified model of particle physics and cosmology, it can naturally integrate the above-mentioned universe ingredients together and really realize connections among them in the universe evolution, of course, this is also fitting to Occam's Razor. Firstly, I put forward to a new extension of the SM, which covers both the SM sector and the dark sector beyond the SM. Secondly, on the basis of the new particle model and the standard cosmological principle, I in sequence research the dynamical evolutions of the relevant energy components of the inflation, the reheating and the current era, among others, I introduce some new ideas and techniques to solve all of the above-mentioned issues successfully and completely. The idea framework of the unified model will be described in the next Section and illustrated by Fig. 3. Lastly, the model numerical results will clearly show the characteristic evolution of each phase, they not only perfectly fit all of the observed data in Eq. (1), but also give many interesting predictions, of course, the model also clarifies and eliminates the ``Hubble tension". In a word, this model can successfully and elegantly account for the origin and evolution of the universe in a unified and integrated way.

 The remainder of this paper is organized as follows. In Section II, I outline the new extension of the SM, and then discuss the neutrino mass, leptogenesis mechanism and dark matter annihilation. I give a complete solution of the slow-roll inflation in Section III. I discuss the reheating evolution and baryogenesis in Section IV. I discusses the current CDM condensation and dark energy genesis, and also eliminating the ``Hubble tension" in Section V. Section VI is a summary of the numerical results of the unified model. Section VII is devoted to conclusions.

\vspace{0.6cm}
\noindent\textbf{II. Particle Model}

\vspace{0.3cm}
 The unified theory is based on the following particle model. I assume that below the GUT scale of $\sim10^{16}$ GeV, the particle contents and symmetries in the universe are showed by Table 1 (where the irrelevant quarks and gauge bosons of the SM are all omitted), all kinds of the notations are explained by the caption.
\begin{table}
 \centering
 \begin{tabular}{|c|c|c|c|c|c|c|c|c|}
  \hline\hline
  &\multicolumn{3}{|c|}{SM (visible sector)} & &\multicolumn{4}{|c|}{BSM (dark sector)} \\\hline
  Fields &$H$ &$l_{\alpha}$ &$e^{-}_{\beta R}$ &$N^{0}_{L}\,,N^{0}_{R}$ &$E^{-}_{L},E^{-}_{R}$ &$\nu^{0}_{\beta R}$ &$\Phi$ &$S\,,\,\phi$ \\\hline
  $SU(2)_{L}\otimes U(1)_{Y}$ &$(2,1)$ &$(2,-1)$ &$(1,-2)$ &$(1,0)$ &$(1,-2)$ &$(1,0)$ &$(2,1)$ &$(1,0)$ \\\hline
  Dark $Z_{2}$ parity &\multicolumn{3}{|c|}{$1$} &$1$ &$1\,,\,-1$ &$-1$ &$-1$ &$1\,,\,-1$ \\\hline
  Dark $Z'_{2}$ parity &\multicolumn{3}{|c|}{$1$} &$1$ &$-1\,,\,-1$ &$1$ &$1$ &$-1\,,\,1$ \\\hline\hline
 \end{tabular}
 \caption{The particle contents and symmetries of the unified model. $H=(H^{+},H^{0})^{T}$ is the scalar doublet, $l_{\alpha}=(\nu^{0}_{\alpha L},e^{-}_{\alpha L})^{T}$ is the left-handed lepton doublet, $(\alpha,\beta=1,2,3)$ are the fermion family indices. $N^{0}$ is a neutral Dirac fermion with a super-heavy mass, $E^{-}$ is a dark charged lepton with a TeV-scale mass, $\nu^{0}_{\alpha L}$ and $\nu^{0}_{\beta R}$ will be combined into the Dirac neutrino with a Sub-eV mass. The dark scalar doublet $\Phi=(\Phi^{+},\Phi^{0})^{T}$ is the super-heavy inflation field, the neutral singlets $S$ and $\phi$ are two real scalar fields. The third row is the quantum numbers under the gauge groups $SU(2)_{L}\otimes U(1)_{Y}$, the last two rows are the dark parities under the dark symmetries $Z_{2}\otimes Z'_{2}$\,. $SU(2)_{L}\otimes U(1)_{Y}$ and $Z_{2}$ are respectively broken by $\langle H\rangle$ and $\langle\phi\rangle$, but $Z'_{2}$ is unbroken due to $\langle S\rangle=0$. The stable $S$ whose mass is about $250$ GeV will become the CDM, furthermore,  the CDM can eventually condense into the dark energy. Note that the global $B-L$ number is always conserved in the model.}
\end{table}
 The SM particles are all in the visible sector, while the particles beyond SM (BSM) are all inhabiting in the dark sector (at least for the current period they are invisible), and they obey the dark symmetry comprised of two discrete groups of $Z_{2}\otimes Z'_{2}$\,. $N^{0}$ is a purely neutral Dirac fermion without any charge but it has a super-heavy mass, it is actually a mediator between the SM sector and the dark one. $E^{-}$ is a dark charged lepton whose mass is $\sim1$ TeV. The dark doublet scalar $\Phi$ is the inflation field whose mass is $\sim10^{10}$ GeV, its dynamical evolution by self-interaction will give rise to the primordial inflation and its decay will lead to reheating and hot big bang of the universe. $S$ and $\phi$ are two real scalars with neutral charges, the former has ``$-1$" parity under $Z'_{2}$\,, while the latter has ``$-1$" parity under $Z_{2}$\,. The electroweak breaking is implemented by $\langle H\rangle\approx174$ GeV as usual, while the dark symmetry $Z_{2}$ is spontaneously broken by $\langle\phi\rangle\sim1$ TeV, but the dark $Z'_{2}$ is always unbroken due to $\langle S\rangle=0$.  After the symmetry breakings, the $E^{-}$ mass is generated by $\langle\phi\rangle$, while $\nu^{0}_{L}$ and $\nu^{0}_{R}$ are combined into the Dirac neutrino, its Sub-eV mass is generated by the following seesaw mechanism. Finally, the unbroken and stable $S$ whose mass is about $250$ GeV will be decoupled after its annihilation is frozen, and then it gradually cools into the current CDM, in the more later stage the CDM becomes super-cool so that it can eventually condenses into the dark energy. Note that the global $B-L$ number is always conserved in the model, in addition, the model symmetry and the fermion arrangement together guarantee that the model is as free-anomaly as the SM. In a word, all of the ingredients required by the universe evolution have fully and perfectly been accommodated in this extension of the SM.

 Based on the particle contents and symmetries in Table 1, the full invariant Lagrangian of the model are
\begin{alignat}{1}
 \mathscr{L}=&\,\overline{N}i\gamma^{\mu}\partial_{\mu}N+\overline{E_{L}}i\gamma^{\mu}D_{\mu}E_{L}
   +\overline{E_{R}}i\gamma^{\mu}D_{\mu}E_{R}+\overline{\nu_{\beta R}}i\gamma^{\mu}\partial_{\mu}\nu_{\beta R} \nonumber\\
 &+(D^{\mu}\Phi)^{\dagger}D_{\mu}\Phi+\frac{1}{2}\partial^{\mu}S\partial_{\mu}S+\frac{1}{2}\partial^{\mu}\phi\partial_{\mu}\phi \nonumber\\
 &+[\,Y^{e}_{\alpha\beta}\overline{l_{\alpha}}\,e_{\beta R}H+Y^{\nu}_{\alpha\beta}\overline{l_{\alpha}}\,\nu_{\beta R}(i\tau_{2}\Phi^{*})
   +y^{l}_{\alpha}\overline{l_{\alpha}}\,N_{R}(i\tau_{2}H^{*})+y^{\nu}_{\beta}\overline{N_{L}}\,\nu_{\beta R}\phi
   -\overline{N_{L}}M_{N}N_{R} \nonumber\\
 &\hspace{0.5cm}+y^{E}\overline{E_{L}}\,E_{R}\phi+y^{e}_{\beta}\overline{E_{L}}\,e_{\beta R}S+h.c.] \nonumber\\
 &+\mu_{0}[\,\phi\,\Phi^{\dagger}H+h.c.]-|H|^{2}(\lambda_{1}\phi^{2}+\lambda_{2}S^{2}+2\lambda_{3}|\Phi|^{2})
   -\phi^{2}(\frac{\lambda_{4}}{2}S^{2}+\lambda_{5}|\Phi|^{2})-\lambda_{6}S^{2}|\Phi|^{2} \nonumber\\
 &-V_{H}-V_{\Phi}-V_{S}-V_{\phi}\,,\nonumber\\
 &V_{H}=-\mu_{H}^{2}|H|^{2}+\lambda_{H}|H|^{4},\hspace{0.5cm}V_{\phi}=-\frac{\mu_{\phi}^{2}}{2}\phi^{2}+\frac{\lambda_{\phi}}{4}\phi^{4},\nonumber\\
 &V_{\Phi}=\mu_{\Phi}^{2}|\Phi|^{2}+\lambda_{\Phi}|\Phi|^{4}+\cdots,\hspace{0.5cm}V_{S}=\frac{\mu_{S}^{2}}{2}S^{2}+\frac{\lambda_{S}}{4}S^{4}+\cdots,
\end{alignat}
 where $D_{\mu}$ is the gauge covariant derivative, $\tau_{2}$ is the second Pauli matrix, and the irrelevant parts of the SM Lagrangian are all omitted. In the Yukawa sector, note that any Majorana-type mass or couplings are all prohibited by the global $B-L$ number conservation. The $N^{0}$ fermion has an inherent mass of $M_{N}\sim10^{8}$ GeV. $[Y^{e}_{\alpha\beta}, Y^{\nu}_{\alpha\beta}, y^{l}_{\alpha}, y^{\nu}_{\beta}, \cdots]$ are all Yukawa coupling parameters, the repeated family indices are summed by default. We can individually rotate the flavor spaces of $l,e_{R},\nu_{R}$ so that $Y^{e}$ becomes a real diagonal matrix (namely the mass eigenstate basis of the charged lepton) and $y^{\nu}$ is also real, then the irremovable complex phases in $Y^{\nu},y^{l},y^{e}$ will become $CP$-violating sources in the lepton sector, this thus provides a condition for the leptogenesis. The scalar sector is somewhat complicated. The triple scalar couplings with $\mu_{0}\sim10^{8}$ GeV is very important because it is a key knot linking all kinds of the following vacua. Since both $M_{N}$ and $\mu_{0}$ are $\sim10^{8}$ GeV, then this indicates that they may result from a special symmetry breaking at this super-high scale. In addition, I assume these couplings between two different scalars $[\lambda_{1},\lambda_{2},\lambda_{3},\cdots]\ll1$, namely they are all very weak and negligible. $V_{H}$ and $V_{\phi}$ have usual self-interacting potentials, but $V_{\Phi}$ and $V_{S}$ have unusual self-interacting potential forms, later we will see the inflationary potential form in Eq. (29), here I only write the quadratic and quartic terms of the series expansions of $V_{\Phi}$ and $V_{S}$ since the $\Phi$ and $S$ masses are only related to these terms, the higher order terms with the dimension being $\geqslant6$ are actually suppressed by even power of the Planck mass. Note that the special potential forms of $V_{\Phi}$ and $V_{S}$ will respectively lead that $\Phi$ has a inflation evolution and $S$ has a condensation evolution. In conclusion, Eq. (2) completely describes all kinds of the interactions among the model particles from the primordial inflation to the present universe.

 The model symmetries are spontaneously broken by the following vacuum structures of the scalar fields,
\begin{alignat}{1}
 &\phi\rightarrow\phi^{0}+v_{\phi}\,,\hspace{0.3cm} H\rightarrow\left[\begin{array}{c}0\\\frac{h^{0}+v_{H}}{\sqrt{2}}\end{array}\right],\hspace{0.3cm}
   \Phi\rightarrow\left[\begin{array}{c}\Phi^{+}\\\Phi^{0}+\frac{v_{\Phi}}{\sqrt{2}}\end{array}\right],\hspace{0.3cm} S\rightarrow S,\nonumber\\
 &v_{\phi}\sim1\:\mathrm{TeV},\hspace{0.5cm} v_{H}\approx246\:\mathrm{GeV},\hspace{0.5cm}
   v_{\Phi}\sim10\:\mathrm{eV},\hspace{0.5cm} \langle S\rangle=0\,,
\end{alignat}
 where these vacuum expectation values are hierarchical and they indicate the sequence of the symmetry breakings. After the vacuum breakings, $\phi^{0}$ and $h^{0}$ become two massive real scalars with neutral charges, $\Phi$ still keeps its original structure due to $v_{\Phi}\ll M_{\Phi}$, and $S$ is always unbroken due to $\langle S\rangle=0$. From the total potential minimum in Eq. (2), we can derive that the above three vacuum expectation values are completely determined by the following system of equations,
\begin{alignat}{1}
 &-\mu_{\phi}^{2}+\lambda_{\phi}v_{\phi}^{2}+\lambda_{1}v_{H}^{2}+\lambda_{5}v_{\Phi}^{2}=\frac{\mu_{0}v_{\Phi}v_{H}}{v_{\phi}}\,,\hspace{0.3cm}
   -\mu_{H}^{2}+\lambda_{H}v_{H}^{2}+\lambda_{1}v_{\phi}^{2}+\lambda_{3}v_{\Phi}^{2}=\frac{\mu_{0}v_{\phi}v_{\Phi}}{v_{H}}\,,\nonumber\\
 &\mu_{\Phi}^{2}+\lambda_{\Phi}v_{\Phi}^{2}+\lambda_{3}v_{H}^{2}+\lambda_{5}v_{\phi}^{2}=\frac{\mu_{0}v_{\phi}v_{H}}{v_{\Phi}}\,,
\end{alignat}
 where all kinds of the parameters are chosen such as $[\lambda_{\phi},\lambda_{H},\lambda_{\Phi}]\sim0.1$, $[\lambda_{1},\lambda_{2},\lambda_{3},\cdots]\ll1$, $\mu_{\phi}\sim v_{\phi}$, $\mu_{H}\sim v_{H}$, $\mu_{\Phi}\sim10^{10}$ GeV and $\mu_{0}\sim10^{8}$ GeV, thus Eq. (4) can guarantee the vacuum stability.

 The symmetry breakings directly give rise to the mass terms of the scalar bosons, after the mass matrix is diagonalized, the mass eigenvalues are given as follows,
\begin{alignat}{1}
 & M^{2}_{h^{0}}\approx2\lambda_{H}v_{H}^{2}\,,\hspace{0.5cm} M^{2}_{\phi^{0}}\approx2\lambda_{\phi}v_{\phi}^{2}\,,\nonumber\\
 &M^{2}_{S}=\mu_{S}^{2}+\lambda_{2}v_{H}^{2}+\lambda_{4}v_{\phi}^{2}+\lambda_{6}v_{\Phi}^{2}\,,\hspace{0.5cm}
   M^{2}_{\Phi}=\frac{\mu_{0}v_{\phi}v_{H}}{v_{\Phi}}\,,
\end{alignat}
 $M_{h^{0}}\approx125$ GeV is exactly the measured mass of the SM Higgs boson, $\phi^{0}$ is the dark neutral boson, the mixing angle between $\phi^{0}$ and $h^{0}$ is $\sim\frac{\lambda_{1}v_{H}}{\lambda_{\phi}v_{\phi}}\ll1$, note that $S$ has no mixing with them. Eq. (5) indicates $M_{\phi^{0}}\sim0.5$ TeV, $M_{S}\approx\mu_{S}\sim250$ GeV, and $M_{\Phi}\approx\mu_{\Phi}\sim10^{10}$ GeV, all of them are suitable values. In fact, we will respectively determine $M_{\Phi}\approx5.05\times10^{10}$ GeV from the inflation solution in Section III and $M_{S}\approx256$ GeV from the CDM condensation in Section V. Later one will see that these mass values have important implications for phenomena of particle physics and cosmology.

 The $e^{-}_{\alpha}$ masses and the $E^{-}$ mass are directly generated by $\langle H\rangle$ and $\langle\phi\rangle$, respectively, note that there is no mixing between  $e^{-}_{\alpha}$ and $E^{-}$ since $S$ is unbroken (namely $\langle S\rangle=0$). By contrast, the $\nu^{0}_{\alpha}$ masses are actually generated by the following effective couplings, which is derived from Eq. (2) by integrating out the super-heavy $\Phi$ and $N$ at the low energy, thus we obtain
\begin{alignat}{1}
 &\mathscr{L}^{eff}_{neutrino}=\overline{l}_{\alpha}\left[Y^{\nu}_{\alpha\beta}\frac{\mu_{0}\phi}{M_{\Phi}^{2}}
   +y^{l}_{\alpha}y^{\nu}_{\beta}\frac{\phi}{M_{N}}\right]\nu_{\beta R}(i\tau_{2}H^{*})\,,\nonumber\\
 \Longrightarrow\; &M_{\nu}=M_{\nu}^{a}+M_{\nu}^{b}
   =-Y^{\nu}_{\alpha\beta}\frac{v_{\Phi}}{\sqrt{2}}-y^{l}_{\alpha}y^{\nu}_{\beta}\frac{v_{\phi}v_{H}}{\sqrt{2}M_{N}}\;\Longrightarrow \sum_{i}m_{\nu_{i}}=\mathrm{Tr}[U^{\nu}_{L}M_{\nu}U_{R}^{\nu\dagger}]\,,\nonumber\\
 &M_{e}=-Y^{e}_{\alpha\beta}\frac{v_{H}}{\sqrt{2}}\,,\hspace{0.5cm} M_{E}=-y^{E}v_{\phi}\,,
\end{alignat}
 where $M_{\nu}$ is diagonalized by these two unitary matrices of $U^{\nu}_{L}$ and $U^{\nu}_{R}$ which respectively rotate $\nu_{\alpha L}$ and $\nu_{\beta R}$. Obviously, this mechanism of generating neutrino mass is a Dirac-type seesaw, which is different from the usual Majorana-type seesaw \cite{22}. In Eq. (6), all kinds of the Yukawa parameters are chosen such as $y^{E}\sim0.5$, $Y^{e}\sim10^{-2}$, $Y^{\nu}\sim10^{-3}$, $y^{l}y^{\nu}\sim10^{-7}$, and $M_{N}\sim10^{8}$ GeV, then we naturally obtain $M_{E}\sim0.5$ TeV, $M_{e}\sim1$ GeV, $M_{\nu}^{a}\sim M_{\nu}^{b}\sim10^{-2}$ eV. Note that $M_{\nu}^{a}$ has three eigenvalues, but $M_{\nu}^{b}$ has only one eigenvalue. Provided $M_{\nu}^{a}\sim0.01$ eV and $M_{\nu}^{b}\sim0.05$ eV, then this may lead to such mass spectrum as $m_{\nu_{1}}\approx0.005<m_{\nu_{2}}\approx0.01<m_{\nu_{3}}\approx0.05$ (eV as unit), thus we naturally explain that $\triangle m_{32}^{2}\approx2.4\times10^{-3}$ $\mathrm{eV^{2}}$ is much larger than $\triangle m_{21}^{2}\approx7.5\times10^{-5}$ $\mathrm{eV^{2}}$, which is exactly required by the experimental data of the neutrino oscillation. Under the flavor basis of real diagonal $Y^{e}$ and real $y^{\nu}$, then $U^{\nu}_{L}$ is identified as the lepton mixing matrix $U_{PMNS}$\,, the complex phases in $Y^{\nu}$ and $y^{l}$ are thus transferred into the $CP$-violating phase in $U_{PMNS}$\,, this $CP$-violating phase is closely related to the following leptogenesis. We can further fit the neutrino mixing angles by choosing a suitable texture of $M_{\nu}$, but here we do not go into it. Based on both the neutrino oscillation experiments and the astrophysics investigations \cite{1}, I will take the suitable $\sum m_{\nu}\approx0.065$ eV as an input parameter of the unified model, see Table 2 in Section VI.

 The dark sector of this model has plentiful phenomena, and they have very important implications for cosmology. Firstly, the $\Phi$ field slow-roll causes the primordial inflation, see Section III. Secondly, after the inflation the $\Phi$ decay brings about the universe reheating and the hot big bang, see Section IV, at the same time, this decay also leads to the matter-antimatter asymmetry by the following leptogenesis mechanism. In the light of Eq. (2), $\Phi$ has two decay modes of $\Phi\rightarrow l^{c}+\nu_{R}$ and $\Phi\rightarrow H+\phi$, in addition, $\Phi\rightarrow l^{c}+\nu_{R}$ and $\Phi^{*}\rightarrow l+\nu_{R}^{c}$ have $CP$ asymmetric decay widths via the interference between the tree diagram amplitude and the one-loop diagram one, as shown in Fig. 1.
\begin{figure}
 \centering
 \includegraphics[totalheight=4cm]{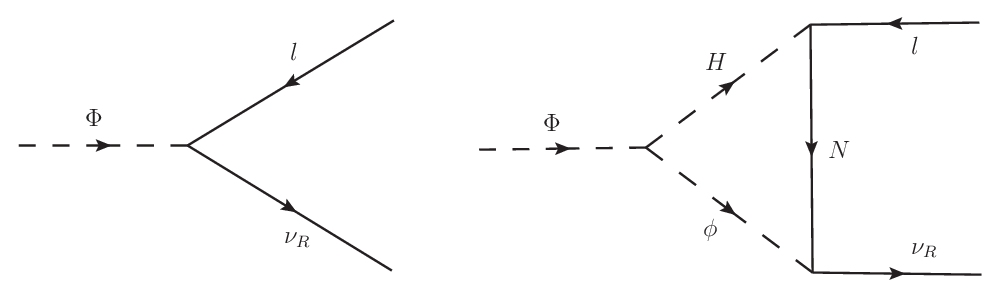}
 \caption{The tree and one-loop diagrams of $\Phi\rightarrow l^{c}+\nu_{R}$\,. The $CP$ asymmetry of this decay can generate asymmetric anti-lepton and asymmetric $\nu_{R}$ with the same number although the net total lepton number is conserved as zero, the former is partly converted into the baryon asymmetry through the SM sphaleron transition, whereas the latter is forever frozen in the dark sector.}
\end{figure}
 The relevant decay width and $CP$ asymmetry are calculated as follows,
\begin{alignat}{1}
 &\Gamma(\Phi\rightarrow l^{c}+\nu_{R})=\frac{M_{\Phi}}{16\pi}\mathrm{Tr}[Y^{\nu\dagger}Y^{\nu}]
   \ll\Gamma(\Phi\rightarrow H+\phi)=\frac{M_{\Phi}}{16\pi}(\frac{\mu_{0}}{M_{\Phi}})^{2},\nonumber\\
 &A_{CP}=\frac{\Gamma(\Phi\rightarrow l^{c}+\nu_{R})-\Gamma(\Phi^{*}\rightarrow l+\nu_{R}^{c})}{\Gamma_{\Phi}} \nonumber\\
 &\hspace{0.8cm}\approx[\frac{1}{2\pi}\frac{M_{N}}{\mu_{0}}ln\frac{M_{N}}{M_{\Phi}}]\,
   \mathrm{Im}[\sum\limits_{\alpha,\beta}Y^{\nu\dagger}_{\beta\alpha}y^{l}_{\alpha}y^{\nu}_{\beta}\,]
   =[\frac{1}{\pi}\frac{M_{N}^{2}}{\mu_{0}^{2}}ln\frac{M_{N}}{M_{\Phi}}]
   \frac{M_{\Phi}^{2}\mathrm{Im}\mathrm{Tr}[M_{\nu}^{a\dagger}M_{\nu}^{b}]}{(v_{\phi}v_{H})^{2}}\,,
\end{alignat}
 where $Y^{\nu}\sim10^{-3}<\frac{\mu_{0}}{M_{\Phi}}\sim10^{-2}$, so the total width $\Gamma_{\Phi}$ is approximately equal to $\Gamma(\Phi\rightarrow H+\phi)$. In Eq. (7), the $CP$-violating sources are from the irremovable complex phases in $Y^{\nu}$ and/or $y^{l}$, accordingly they are contained in $M_{\nu}^{a}$ and/or $M_{\nu}^{b}$, therefore the $CP$ violation in the leptogenesis is closely related to the $CP$ violation of $U_{PMNS}$ measured by the neutrino experiments \cite{23}. Since $\frac{M_{N}}{\mu_{0}}\sim1$ and $\frac{M_{N}}{M_{\Phi}}\sim10^{-2}$, the factor terms of the two square brackets are all $\sim1$. Provided $Y^{\nu}\sim10^{-3}$ and $y^{l}y^{\nu}\sim10^{-7}$ as before, then we can naturally obtain $A_{CP}\sim10^{-10}$, in other terms, since there are $M_{\Phi}\sim10^{10}$ GeV, $M_{\nu}\sim10^{-10}$ GeV and $v_{\phi}v_{H}\sim10^{5}$ $\mathrm{GeV}^{2}$, then we certainly obtain $A_{CP}\sim10^{-10}$. This value is very vital for the baryon asymmetry, see the following Eq. (37). In addition, by a simple calculation one can prove that $\Gamma(\Phi\rightarrow l^{c}+\nu_{R})$ is smaller than the universe expansion rate at the temperature of $T=M_{\Phi}\approx5.05\times10^{10}$ GeV, so the decay in Fig. 1 is really an out-of-equilibrium process. Note that the dilute process of $l^{c}+\nu_{R}\rightarrow\phi+H$ via the t-channel $N$ mediation is invalid at any temperature because $\frac{|y^{l}y^{\nu}|^{2}}{M_{N}}\ll\frac{1}{M_{Pl}}$ ($M_{Pl}$ is the Planck mass) guarantees that its reaction rate is always severely out-of-equilibrium. As a result, the $CP$ asymmetry in Eq. (7) can generate asymmetric anti-lepton and asymmetric $\nu_{R}$ with the same number although the net total lepton number is conserved as zero, the former will be partly converted into the baryon asymmetry through the sphaleron transition in the SM sector \cite{24}, see Section IV, whereas the latter is forever frozen in the dark sector. Finally, I should stress that the amount of the matter-antimatter asymmetry is closely related to these fundamental quantities of $M_{\Phi}$, $M_{\nu}$, $v_{\phi}$ and $v_{H}$ by Eq. (7) and the following Eq. (37), this is rightly a characteristic of the unified model.

 The hot evolution of the dark sector is parallel to that of the SM sector. After reheating all kinds of the dark particles are thermally produced by those interactions in Eq. (2), however, these heavy dark particles of $N^{0},E^{-},\phi^{0}$ are early depleted by their decays. Since $S$ is unbroken (namely $\langle S\rangle=0$), the dark parity of $Z'_{2}$ is always conserved, in addition, provided $M_{S}<M_{E}$, then these guarantee that $S$ is a stable dark particle without any decay. Nevertheless, a pair of $S$ can annihilate into $S+S\rightarrow e_{\alpha R}+e^{c}_{\beta R}$ via the dark $E^{-}$ mediation, as shown Fig. 2.
\begin{figure}
 \centering
 \includegraphics[totalheight=4cm]{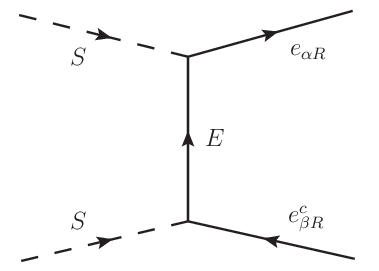}
 \caption{The annihilation of $S+S\rightarrow e_{\alpha R}+e^{c}_{\beta R}$ via the dark $E^{-}$ mediation, which leads that the residual $S$ particles will become the current CDM and eventually condensate into the dark energy. This process can be searched as the direct and indirect detections for the CDM.}
\end{figure}
  When the universe cools to the freeze-out temperature, the annihilation process is terminated and the $S$ particle are decoupled from the $e_{\alpha R}$ leptons, the residual $S$ particles thus become the current CDM. As the universe temperature is more and more cooling, the supercool CDM can eventually condensate into the dark energy, see Section V. The thermally averaged annihilation cross-section and the freeze-out temperature are simply calculated by the following relations,
\begin{alignat}{1}
 &\langle\sigma v_{r}\rangle_{T_{f}}=\langle c_{1}+c_{2}v_{r}^{2}+c_{3}v_{r}^{4}+\cdots\rangle_{T_{f}}
   \approx c_{1}+c_{2}\frac{6T_{f}}{M_{S}}\,,\hspace{0.3cm}
   c_{1}=0\,,\hspace{0.3cm} c_{2}=\frac{M^{2}_{S}}{48\pi M^{4}_{E}}\sum\limits_{\alpha,\beta}|y^{e}_{\alpha}y^{e}_{\beta}|^{2}\,,\nonumber\\
 &\langle\sigma v_{r}\rangle_{T_{f}} n_{S}(T_{f})=H(T_{f})=\frac{1.66\sqrt{g_{*}(T_{f})}\,T_{f}^{2}}{M_{Pl}}\,,\hspace{0.3cm}
   n_{S}(T_{f})=T_{f}^{3}[\frac{M_{S}}{2\pi T_{f}}]^{\frac{3}{2}}e^{-\frac{M_{S}}{T_{f}}},\nonumber\\
 &\Longrightarrow \frac{M_{S}}{T_{f}}\approx 20+\frac{1}{2}\,ln\frac{\frac{M_{S}}{T_{f}}}{g_{*}(T_{f})}
   +ln\frac{M_{S}\langle\sigma v_{r}\rangle_{T_{f}}}{10^{-9}\,\mathrm{GeV}^{-1}}\,,
\end{alignat}
 where $v_{r}=2\sqrt{1-\frac{4M^{2}_{S}}{s}}$ is the relative velocity (where s is the squared center-of-mass energy), $M_{Pl}$ is the Planck mass, $g_{*}(T_{f})=91.5$ is the effective number of relativistic degrees of freedom at $T_{f}\approx10.1$ GeV. Provided $\sum\limits_{\alpha,\beta}|y^{e}_{\alpha}y^{e}_{\beta}|^{2}\approx1$, $M_{E}\approx0.5$ TeV and $M_{S}\approx256$ GeV, then we can calculate out $\langle\sigma v_{r}\rangle_{T_{f}}\approx1.64\times10^{-9}$ $\mathrm{GeV}^{-2}$ and $\frac{M_{S}}{T_{f}}\approx25.4$, so there is $T_{f}\approx10.1$ GeV. These values are very vital for the current density budget of the CDM and dark energy, which will be discussed in Section V. Finally, we can make use of the elastic or non-elastic scattering of $S+e_{\alpha R}\rightarrow S+e_{\beta R}$ to detect the CDM, which is namely rotating Fig. 2 by 90 degree.

 On the basis of the above-mentioned particle model, we can describe the idea framework of the universe origin and evolution by the sketch shown as Fig. 3.
\begin{figure}
 \centering
 \includegraphics[totalheight=7cm]{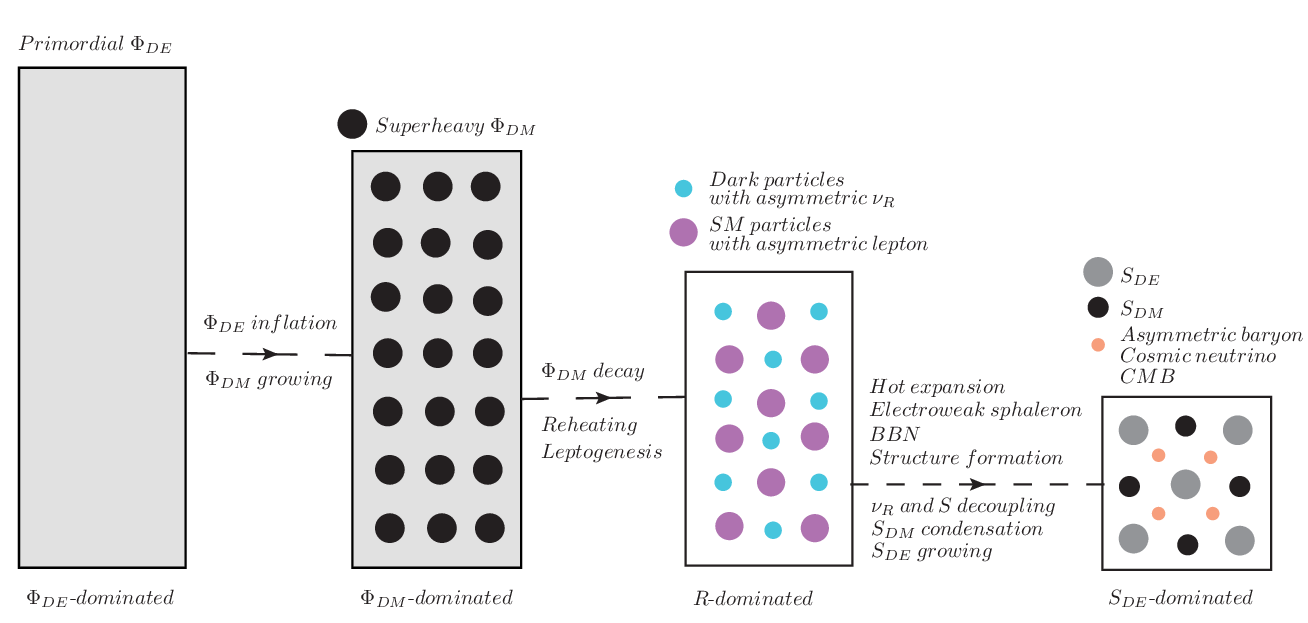}
 \caption{The sketch of the universe origin and evolution described by the unified model. The universe energy is step by step released and reduced from the primordial dark energy $\Phi_{DE}$ to the current $S_{DE}$-dominated energy budget, the whole evolution process is analogous to a cascade of hydropower stations, there is not the so-called ``cosmological constant energy" in the model.}
\end{figure}
 In sequence, the universe passed through the primordial inflation, the followed reheating, the early hot expansion, the transformation from the radiation-dominated to the matter-dominated, the supercool CDM condensing into the dark energy, and the present DE-dominated universe. The primordial inflation is implemented by the $\Phi$ field slowly rolling. $\Phi$ has the two physical phase states or energy forms of $\Phi_{DE}$ and $\Phi_{DM}$ because it has special nature as an unusual matter. $\Phi_{DE}$ is an inert condensed phase state with a negative pressure, it has no kinetic energy and can not take part in couplings to the other particles, whereas $\Phi_{DM}$ is an excited massive particle phase state with a vanishing pressure, it has kinetic energy and can interact with the other particles, see the following Eq. (10). Inappropriately, the relationship between $\Phi_{DE}$ and $\Phi_{DM}$ is analogous to ice and vapour, which are merely the two different physical phase states of the same material. The same physical meanings and explanation also apply to the $S$ field, namely it has also the two physical phase states or energy forms of $S_{DE}$ and $S_{DM}$, see Eq. (39). In the following Sections, we will see that the physical essence of the slow-roll inflation is an evolution that the super-heavy dark matter $\Phi_{DM}$ is slowly growing from the primordial dark energy $\Phi_{DE}$, namely $\Phi_{DE}$ is gradually converting into $\Phi_{DM}$. After the inflation is finished, the $\Phi_{DM}$ decay is responsible for the reheating and the leptogenesis. When the hot bath is formed and the radiation begins to dominate the universe, the asymmetric anti-lepton and the asymmetric right-handed neutrino have been generated equally, but they are isolated in the visible sector and the dark sector, respectively. In the hot expansion stage, the asymmetric anti-lepton can be partly converted into the baryon asymmetry through the SM sphaleron transition in the visible sector, whereas the asymmetric right-handed neutrino is forever frozen out in the dark sector. The followed evolution in the visible sector is well-known. In the dark sector, $S_{DM}$ is decoupled from the hot bath below the $T_{f}$ temperature, as the universe temperature cooling, $S_{DM}$ becomes the current CDM. In the very later stage, the temperature is more and more approaching to absolute zero, the kinetic energy of $S_{DM}$ is more and more exhausted, thus the supercool $S_{DM}$ can eventually condense into $S_{DE}$ which is namely the current dark energy, this condensation is in essence an evolution that $S_{DE}$ is slowly growing from $S_{DM}$ or $S_{DM}$ is gradually converting into $S_{DE}$, therefore, the current condensation is essentially a reverse process of the primordial inflation. Although there is a great difference about 106 orders of magnitude between the density of the primordial dark energy $\Phi_{DE}$ and the density of the current dark energy $S_{DE}$, the universe energy is step by step released and reduced through a cascade of the evolutions, this is analogous to a cascade of hydropower stations at the Changjiang River, by which a huge drop of water potential is step by step converted into electrical energy, therefore there is not naturally the so-called ``cosmological constant energy" in the model. Finally, I emphasize that all of these assumptions of the unified model are moderate, reasonable and consistent, by which we can successfully and completely account for the universe origin and evolution.

\vspace{0.6cm}
\noindent\textbf{III. Primordial Inflation}

\vspace{0.3cm}
 The dynamic evolution of the primordial inflation is described by what follows. According to the standard paradigm \cite{25}, the inflation field $\Phi$ is considered as spatially uniform distribution, but there are very small fluctuations, which will become sources of the structure formation. Under the flat FLRW metric, namely $ g_{\mu\nu}=\mathrm{Diag}(1,-a^{2},-a^{2},-a^{2})$ where $a(t)$ is the scale factor of the universe expansion, the energy density and pressure of $\Phi$ are given by its energy-momentum tensor as follows,
\begin{alignat}{1}
 &\mathscr{L}_{\Phi}=g^{\mu\nu}\partial_{\mu}\Phi^{\dagger}\partial_{\nu}\Phi-V_{\Phi},\hspace{0.5cm}
   T^{\mu}_{\:\nu}(\Phi)=2g^{\mu\beta}\partial_{\beta}\Phi^{\dagger}\partial_{\nu}\Phi-\delta^{\mu}_{\,\,\nu}\mathscr{L}_{\Phi},\nonumber\\
 \Longrightarrow\; &T^{0}_{\:\:0}=\rho_{\Phi}=|\dot{\Phi}|^{2}+V_{\Phi},\hspace{0.5cm}
   -\frac{1}{3}\delta^{i}_{\,j}T^{j}_{\:\:i}=P_{\Phi}=|\dot{\Phi}|^{2}-V_{\Phi},
\end{alignat}
 where $\mathscr{L}_{\Phi}$ is the Lagrangian of pure $\Phi$, $V_{\Phi}=V(\Phi^{\dagger}\Phi)=V(|\Phi|^{2})$ is its self-interacting potential energy, $\dot{\Phi}=\frac{d\Phi}{dt}$ and $|\dot{\Phi}|^{2}=\dot{\Phi}^{\dagger}\dot{\Phi}=\dot{\Phi}^{+}\dot{\Phi}^{-}+\dot{\Phi}^{0*}\dot{\Phi}^{0}$ is the kinetic energy of $\Phi$. Obviously, the potential energy and the kinetic energy together determine $\rho_{\Phi}$ and $P_{\Phi}$, and vice versa. Note that once the function form of $V(|\Phi|^{2})$ is provided, $\dot{\Phi}$ is not independent, in principle it should be solved out by the equation of motion of the $\Phi$ field. However, $\rho_{\Phi}$ and $P_{\Phi}$ are all super-high values in the inflation period.

 I now introduce the dark energy $\Phi_{DE}$ and the dark matter $\Phi_{DM}$, they are merely two energy forms or phase states of the $\Phi$ field, each of them has own density and pressure, which are determined by the following relations,
\begin{alignat}{1}
 &P_{\Phi_{DE}}=-\rho_{\Phi_{DE}},\hspace{0.3cm} P_{\Phi_{DM}}=0\,,\hspace{0.3cm}
  \rho_{\Phi_{DE}}+\rho_{\Phi_{DM}}=\rho_{\Phi},\hspace{0.3cm} P_{\Phi_{DE}}+P_{\Phi_{DM}}=P_{\Phi}=w_{\Phi}\rho_{\Phi},\nonumber\\
 \Longrightarrow\; &\rho_{\Phi_{DE}}=-w_{\Phi}\rho_{\Phi}=\frac{-2w_{\Phi}}{1-w_{\Phi}}V_{\Phi},\hspace{0.3cm}
  \rho_{\Phi_{DM}}=(1+w_{\Phi})\rho_{\Phi}=2|\dot{\Phi}|^{2}=|\dot{\Phi}|^{2}+\frac{1+w_{\Phi}}{1-w_{\Phi}}V_{\Phi},
\end{alignat}
 where I employ Eq. (9). $w_{\Phi}$ is a parameter-of-state varying with the time, which relates the total pressure to the total energy density, there is generally $-1\leqslant w_{\Phi}(t)\leqslant0$, $\Phi$ is purely $\Phi_{DE}$ when $w_{\Phi}=-1$, while $\Phi$ entirely becomes $\Phi_{DM}$ when $w_{\Phi}=0$. $\Phi_{DE}$ is an inert condensed state with a negative pressure, it has only potential energy without kinetic energy, so $\Phi_{DE}$ can not interact with the other particles, in contrast, $\Phi_{DM}$ is an excited massive particle state with a vanishing pressure, it carries both kinetic energy and potential energy, and the both are always equal to each other, so $\Phi_{DM}$ can take part in couplings to the other particles. In short, Eq. (10) explicitly shows the inherent relations among all kinds of the energy forms of the $\Phi$ field, the physical implications of $\Phi_{DE}$ and $\Phi_{DM}$ will be further clear in the following context. 

 At the beginning of the inflation, the $\Phi$ field is purely in the $\Phi_{DE}$ phase state, then $\Phi_{DM}$ is very slowly growing from $\Phi_{DE}$, as $\Phi_{DM}$ is more and more generated and $\Phi_{DE}$ is more and more depleted, thus $\Phi_{DE}$ gradually converts into $\Phi_{DM}$, at the end the $\Phi$ field entirely becomes the $\Phi_{DM}$ phase state, this process is namely so-called slow-roll inflation. The dynamics of the inflationary evolution are collectively determined by the Friedmann equation, the $\rho_{\Phi}$ continuity equation and the $\Phi_{DM}$ growth equation, which are respectively
\begin{alignat}{1}
 &\rho_{\Phi}=\rho_{\Phi_{DE}}+\rho_{\Phi_{DM}}=3\tilde{M}^{2}_{p}H^{2},\nonumber\\
 &\dot{\rho}_{\Phi}+3H\rho_{\Phi}(1+w_{\Phi})=0\:\Longrightarrow -\dot{\rho}_{\Phi_{DE}}=\dot{\rho}_{\Phi_{DM}}+3H\rho_{\Phi_{DM}},\nonumber\\
 &\dot{\rho}_{\Phi_{DM}}=-2\eta(t)H\rho_{\Phi_{DM}},
\end{alignat}
 where I employ Eq. (10). $\tilde{M}_{p}=\frac{1}{\sqrt{8\pi G}}\approx2.43\times10^{18}$ GeV is the reduced Planck mass, $H(t)=\frac{\dot{a}(t)}{a(t)}$ is the universe expansion rate, the proportional parameter $-\eta(t)>0$ controls the $\Phi_{DM}$ growth rate, in fact $\eta$ is identified as one of the slow-roll parameters defined below. In Eq. (11), once the evolution of $\eta(t)$ is specified, then the system of equations is closed, from which we can solve all the evolutions of $\rho_{\Phi_{DM}}$, $\rho_{\Phi_{DE}}$, $\rho_{\Phi}$ and $H$. The above continuity equation indicates that the $\rho_{\Phi_{DM}}$ growth in the comoving volume is purely from the $\rho_{\Phi_{DE}}$ reduction, therefore the primordial inflation is indeed the slow process of $\Phi_{DM}$ growing from $\Phi_{DE}$\,.

 From Eqs. (10) and (11), we can easily derive
\begin{alignat}{1}
 &\eta(t)=-\frac{dln\rho_{\Phi_{DM}}}{2Hdt}=-\frac{dln|\dot{\Phi}|}{Hdt}\,,\hspace{0.5cm}
   \epsilon(t)=-\frac{dln\rho_{\Phi}}{2Hdt}=-\frac{\dot{H}}{H^{2}}=\frac{3(1+w_{\Phi})}{2}\,,\\
 &-1=w_{\Phi}(0)\leqslant w_{\Phi}(t)\leqslant w_{\Phi}(t_{inf})=0\,,\hspace{0.5cm}
   0=\epsilon(0)\leqslant\epsilon(t)\leqslant\epsilon(t_{inf})=\frac{3}{2}\,,
\end{alignat}
 where $\eta$ and $\epsilon$ are two slow-roll parameters defined as usual, they respectively characterize the logarithmic varying rate of $\rho_{\Phi_{DM}}$ and that of $\rho_{\Phi}$, of course, $\eta$ and $\epsilon$ themselves are also varying in the inflationary period, or else the inflation will continue on without termination. Eq. (13) gives the inflationary boundary condition, hereinafter we take $t=0$ as the time of inflation begin and use the ``inf" subscript to denote the time of inflation finish. In addition, we can obtain the expansion acceleration equation,
\ba
 \frac{\ddot{a}}{a}=(1-\epsilon)H^{2}=-\frac{1+3w_{\Phi}}{2}H^{2}.
\ea
 Eq. (14) shows that the accelerating or decelerating expansion only depends on the value of $\epsilon$ or $w_{\Phi}$. When $0\leqslant\epsilon\leqslant1$ (or $-1\leqslant  w_{\Phi}\leqslant-\frac{1}{3}$), there is $\ddot{a}(t)\geqslant0$, the early stage is the $\Phi_{DE}$-dominated universe, when $1<\epsilon\leqslant\frac{3}{2}$ (or $-\frac{1}{3}<w_{\Phi}\leqslant0$), there is $\ddot{a}(t)<0$, the later stage is the $\Phi_{DM}$-dominated universe. However, $\dot{H}\leqslant0$ indicates that the expansion rate and the total energy density are always decreased in the inflation period.

 Put Eq. (10) and Eq. (12) together, we can use $\epsilon$ to express the ratios of all kinds of the energy components to the total $\rho_{\Phi}$ as follows,
\ba
 \frac{\rho_{\Phi_{DE}}}{\rho_{\Phi}}=(1-\frac{2\epsilon}{3}),\hspace{0.3cm}\frac{\rho_{\Phi_{DM}}}{\rho_{\Phi}}=\frac{2\epsilon}{3}\,,\hspace{0.3cm} 
 \frac{|\dot{\Phi}|^{2}}{\rho_{\Phi}}=\frac{\epsilon}{3}\,,\hspace{0.3cm} \frac{V_{\Phi}}{\rho_{\Phi}}=(1-\frac{\epsilon}{3}).
\ea
 Eqs. (11), (12) and (15) together make up the fundamental equations of the inflationary evolutions, and Eq. (13) is the boundary condition. If we can provide the evolution of any one of the nine inflationary quantities, $H$, $\rho_{\Phi}$, $\rho_{\Phi_{DE}}$, $\rho_{\Phi_{DM}}$, $|\dot{\Phi}|^{2}$, $V_{\Phi}$, $\epsilon$, $\eta$, $w_{\Phi}$, in principle, then the solutions of the other inflationary quantities will completely be determined by this system of equations. In what follows, we will find the solution for the inflation problem by a special technique.

 One of the inflationary features is that the universe size or the scale factor expands about $10^{25}$ times in an extremely short duration, therefore we use the e-fold number to characterize the inflationary time span instead of the scale factor, it is defined as follows,
\begin{alignat}{1}
 &N(t)=ln\frac{a(t_{inf})}{a(t)}=\int^{t_{inf}}_{t} H(t')dt'\:\Longrightarrow \dot{N}(t)=-H(t),\\
 &0=a(0)\leqslant a(t)\leqslant a(t_{inf}),\hspace{0.3cm} +\infty=N(0)\geqslant N(t)\geqslant N(t_{inf})=0,\nonumber
\end{alignat}
 where the starting point of the inflation is set as $a(0)=0$ and $N(0)=+\infty$. Eq. (16) now acts as the role of $\frac{\dot{a}}{a}=H$ since $N(t)$ replaces $a(t)$ as the time scale, it will frequently be used in the following formula derivations.

 By use of Eqs. (11), (12), (15) and (16), we can also order by order give the slow-roll parameters by the total energy density $\rho_{\Phi}$ and its derivative as follows,
\begin{alignat}{1}
 &\rho'_{\Phi}=\frac{d\rho_{\Phi}}{dN}=3\rho_{\Phi_{DM}},\hspace{0.3cm}
   \rho''_{\Phi}=\frac{d^{2}\rho_{\Phi}}{dN^{2}}=3\rho'_{\Phi_{DM}},\hspace{0.3cm}
   \rho'''_{\Phi}=\frac{d^{3}\rho_{\Phi}}{dN^{3}}=3\rho''_{\Phi_{DM}},\:\ldots\nonumber\\
 &\epsilon=\frac{dln\rho_{\Phi}}{2dN}\,,\hspace{0.3cm}\eta=\frac{dln\rho'_{\Phi}}{2dN}=\frac{dln\rho_{\Phi_{DM}}}{2dN}\,,\hspace{0.3cm}
   \theta=\frac{dln|\rho''_{\Phi}|}{2dN}=\frac{dln|\rho'_{\Phi_{DM}}|}{2dN}\,,\hspace{0.3cm}\delta=\frac{dln|\rho'''_{\Phi}|}{2dN}\,,\:\ldots\\
 \Longrightarrow\; &\frac{dln\epsilon}{2dN}=\eta-\epsilon\,,\hspace{0.5cm} \frac{dln|\eta|}{2dN}=\theta-\eta\,,\hspace{0.5cm}
   \frac{dln|\theta|}{2dN}=\delta-\theta\,,\:\ldots\nonumber\\
 &\frac{dln\rho_{\Phi_{DE}}}{2dN}=\epsilon[\frac{3-2\eta}{3-2\epsilon}]\,,\hspace{0.5cm} 
   \frac{dlnV_{\Phi}}{2dN}=\epsilon[\frac{3-\eta}{3-\epsilon}]\,,
\end{alignat}
 where hereinafter the ``$'$" superscript denotes a derivative with regard to $N$. These slow-roll parameters in Eq. (17) are closely related to the observable quantities of the inflation, for instance, $\epsilon$ is related to the tensor-to-scalar ratio, $\eta$ is relevant to the scalar spectral index, $\theta$ is involved in the running of the spectral index, see the following Eq. (27). Eq. (18) are very useful relations for the following derivations. In short, finding the correct solutions of these slow-roll parameters is a key of solving the inflation problem.

 From the previous fundamental equations and Eq. (16), we can also derive the equation of motion of the $\Phi$ field and its formal solution as follows,
\begin{alignat}{1}
 &\ddot{\Phi}+3H\dot{\Phi}+\Phi\frac{dV_{\Phi}}{d|\Phi|^{2}}=0\:\Longrightarrow
   \frac{\Phi''}{3-\epsilon}-\Phi'+\Phi\frac{d\,lnV_{\Phi}}{d\,[\frac{|\Phi|}{\tilde{M}_{p}}]^{2}}=0\,,\nonumber\\
 &\Longrightarrow \Phi_{1}(N)=\Phi_{2}(N)=\Phi_{3}(N)=\Phi_{4}(N).\\
 &\varphi=\sqrt{2}\,|\Phi|=\sqrt{\sum\limits_{i}\Phi_{i}^{2}}\,,\hspace{0.5cm} \sqrt{2}\,|\Phi'|=\sqrt{\sum\limits_{i}\Phi_{i}'^{2}}\,,\nonumber\\
 &\Longrightarrow\frac{d\varphi}{dN}=\sqrt{2}\,\frac{d|\Phi|}{dN}=\sqrt{4}\,\frac{d|\Phi_{i}|}{dN}=\pm\sqrt{4}\,|\Phi'_{i}|
   =\pm\sqrt{2}\,|\Phi'|=\pm\tilde{M}_{p}\sqrt{2\epsilon}\,,\nonumber\\
 &\Longrightarrow\frac{\varphi(N)}{\sqrt{2}\tilde{M}_{p}}=\frac{|\Phi(N)|}{\tilde{M}_{p}}
   =\frac{|\Phi(\infty)|}{\tilde{M}_{p}}-\int_{\infty}^{N}\sqrt{\epsilon(N')}dN'=\int_{N}^{\infty}\sqrt{\epsilon(N')}dN',
\end{alignat}
 where $\Phi'=\frac{d\Phi}{dN}$ and  $\Phi_{i}'=\frac{d\Phi_{i}}{dN}$, $\Phi_{i}$ (i=1,2,3,4) are four real degree of freedoms of $\Phi$. Obviously the solutions of $\Phi_{i}$ are degenerate because $V(|\Phi|^{2})$ is fully symmetric with regard to each of $\Phi_{i}$. Note that when $\frac{dV_{\Phi}}{d|\Phi|^{2}}>0$, it has namely the meaning of $M_{\Phi}^{2}$. In Eq. (20), I introduce an auxiliary field $\varphi$ in order to get rid of the multiple components of $\Phi$, furthermore, I assume that $\varphi$ (namely $|\Phi|$) is gradually increased from the initial $\varphi(\infty)=0$ in the slow-roll process, therefore there is $\dot{\varphi}>0$ (namely $\varphi'<0$), the physical meanings for this choice will be seen later. Once $\epsilon(N)$ is provided, we can immediately calculate out the $\varphi$ (namely $|\Phi|$) evolution.

 According to traditional procedure, if the inflationary potential $V(|\Phi|^{2})$ is provided, then the conventional slow-roll parameters are given by $V_{\Phi}$ as follows,
\begin{alignat}{1}
 &\epsilon_{V}=\frac{\tilde{M}^{2}_{p}}{2}[\frac{dV_{\Phi}}{V_{\Phi}d\varphi}]^{2}=\epsilon\,[\frac{3-\eta}{3-\epsilon}]^{2}
   \xrightarrow{\epsilon,\,\eta\ll1}\epsilon\,,\nonumber\\
 &\eta_{V}=\tilde{M}^{2}_{p}[\frac{d^{2}V_{\Phi}}{V_{\Phi}d\varphi^{2}}]=\frac{(3-\eta)(\epsilon+\eta)-\eta'}{3-\epsilon}
   \xrightarrow{\epsilon,\,\eta,\,\theta\ll1}\epsilon+\eta-\frac{\eta'}{3}\,,\nonumber\\
 &\xi^{2}_{V}=\tilde{M}^{4}_{p}[\frac{dV_{\Phi}}{V_{\Phi}d\varphi}][\frac{d^{3}V_{\Phi}}{V_{\Phi}d\varphi^{3}}]
                     =[\frac{3-\eta}{3-\epsilon}]^{2}\,[4\epsilon\eta+\frac{\eta'(3-3\epsilon+2\eta-2\theta)-2\eta\theta'}{3-\eta}]\nonumber\\
 &\hspace{2.8cm}\xrightarrow{\epsilon,\,\eta,\,\theta,\,\delta\ll1}4\epsilon\eta+\eta'-\frac{2\eta\theta'}{3}\,,
\end{alignat}
 where $\eta'=\frac{d\eta}{dN}$ and $\theta'=\frac{d\theta}{dN}$, and I employ the foregoing equations to derive the above relations. These two sets of slow-roll parameters are now related by Eq. (21). It should however be stressed that the above approximations are held only when $\epsilon,\eta,\theta,\delta\ll1$, this case is only in the early and middle phases of the inflation, when the inflation is close to its end, the slow-roll parameters actually become $\sim1$, thus these approximations are all invalid.

 We will later see that the evolution at the end of inflation is very important. We can make Taylor expansion of $V_{\Phi}(\varphi)$ around $\varphi(t_{inf})=\varphi_{f}$, thus obtain the following results,
\begin{alignat}{1}
 &V_{\Phi}(\varphi)=V_{\Phi}(\varphi)|_{\varphi_{f}}+\frac{dV_{\Phi}(\varphi)}{d\varphi}|_{\varphi_{f}}(\varphi-\varphi_{f})
   +\frac{1}{2}\frac{d^{2}V_{\Phi}(\varphi)}{d\varphi^{2}}|_{\varphi_{f}}(\varphi-\varphi_{f})^{2}+\cdots\,,\nonumber\\
 &\frac{\varphi(t_{inf})}{\sqrt{2}\tilde{M}_{p}}=\frac{|\Phi(t_{inf})|}{\tilde{M}_{p}}=\int_{0}^{\infty}\sqrt{\epsilon(N)}dN,\\
 &M^{2}_{\Phi}=\frac{d^{2}V_{\Phi}(\varphi)}{d\varphi^{2}}|_{\varphi_{f}}=[(3-\epsilon)\eta_{V}H^{2}]_{t_{inf}}\:\Longrightarrow
   M_{\Phi}=[\sqrt{(3-\epsilon)\eta_{V}}H\,]_{N=0}\,,
\end{alignat}
 where I employ Eq. (20) and Eq. (21). From the above equations, we can calculate out the field value and the $\Phi$ mass at the inflation end. Note that $M_{\Phi}$ may be identified as the meaning of the $\Phi$ mass only when $\eta_{V}$ becomes positive, in the following Fig. 4, we will see how $M_{\Phi}$ is gradually generated from nothing as $\eta_{V}$ evolving from negative to positive, namely $M_{\Phi}$ is generated by a special mechanism of the inflation evolution. Later we will calculate out $|\Phi(t_{inf})|/\tilde{M}_{p}\approx4.98$ and $M_{\Phi}\approx5.05\times10^{10}$ GeV, these are all very reasonable values.

 A traditional and usual technique of solving the inflation problem is as the following procedure. Firstly, one has to design or guess a function form of $V(|\Phi|^{2})$ (namely $V(\varphi)$). Secondly, one puts $V_{\Phi}$ into Eq. (19), and then ignores the $\epsilon$ factor because there is $\epsilon\ll1$ in the most of the inflation duration, thus one can solve the differential equation of $\Phi$ to find a solution of $\Phi(N)$. Thirdly, from Eq. (20) one can obtain $\epsilon(N)$ by making a derivative of $\Phi(N)$, and further one can calculate $\eta(N)$ and $\theta(N)$ by Eq. (18). Lastly, since $V_{\Phi}(N)$ and $\epsilon(N)$ have been obtained, one can then work out $\rho_{\Phi}$, $\rho_{\Phi_{DE}}$, $\rho_{\Phi_{DM}}$ and $|\dot{\Phi}|^{2}$ by Eq. (15), By this time the inflationary evolutions are completely solved out. Nevertheless, this procedure has two serious shortcomings. i) In the later phase of the inflation $\epsilon$ is actually $\sim1$ rather than $\ll1$, therefore one neglecting $\epsilon$ in Eq. (19) will lead to a non-rigorous and incomplete inflationary solution, in particular, this has great effect on the inflation termination and the followed reheating. ii) it is very difficult to fit precisely all of the inflationary data in Eq. (1) by this technique, in fact, a desirable inflationary potential is not be found as yet although the countless endeavours have been made. In conclusion, to solve the inflation evolution reliably and completely, we have to find a new approach.

 In the system of equations of the inflationary evolution, all kinds of the inflationary quantities have an equal status at least in mathematical sense, once one of them is provided, the rest of them can all be solved. Therefore we can flexibly choose the parameter $\eta$ as the starting point instead of the potential $V_{\Phi}$. In brief, we can employ the following procedure. Firstly, we need design or guess a function form of $\eta(N)$ such as the following Eq. (24), this amounts to specifying directly the law of the $\Phi_{DM}$ growth in Eq. (11), thereby Eq. (11) becomes a closed the system of equations, in principle we can solve out $\rho_{\Phi_{DM}}$, $\rho_{\Phi_{DE}}$, $\rho_{\Phi}$ and $H$. Secondly, we can further figure out $|\dot{\Phi}|^{2}$, $V_{\Phi}$, $\epsilon$ and $w_{\Phi}$ by Eqs. (15) and (12). Thirdly,  we can calculate out $\varphi(N)$ by Eq. (20), and then derive the numerical solution of $V(\varphi)$ from $V(N)$ and $\varphi(N)$. Lastly, we will find out the analysis function of $V(\varphi)$ by fitting its numerical solution, thus the inflationary potential is reversely solved out rather than provided early. By this time all of the inflationary problems are completely solved out. Obviously, this technique overcome the shortcomings of the traditional technique, it can give exact and complete solution of the inflation evolution. Of course, whatever technical means is employed, the only criterion is that it is able to fit all of the inflationary data correctly and completely.

 After careful analyses and calculations, I give an appropriate function form of $\eta(N)$ as follows,
\begin{alignat}{1}
 &\eta(N)=(\frac{3}{2}+C)\frac{1-\beta N}{1+N}e^{-\alpha\frac{N^{2}}{N_{*}^{2}}}-C,\\
 \Longrightarrow\; &\eta(0)=\frac{3}{2}\,,\hspace{0.3cm} \eta(\frac{1}{\beta})=0,\hspace{0.3cm} \eta(+\infty)=-C,\hspace{0.3cm} \eta(-1)=+\infty,\hspace{0.3cm} \eta'(0)=-(\frac{3}{2}+C)(\beta+1),\nonumber
\end{alignat}
 where $\alpha=3.22$, $\beta=0.3$, $C=0.001$ are three independent parameters of the inflation model. $N_{*}$ is corresponding to the inflationary e-fold number when the pivot scale of $k_{*}=0.05$ $\mathrm{Mpc}^{-1}$ exits from the horizon, it will be calculated as $N_{*}\approx52.43$ by the following Eq. (28). The values of $\alpha$ and $\beta$ are determined by fitting the data of the scalar spectral index and spectral index running, while the $C$ value is relatively free fixed because there is only an upper bound for the tensor-to-scalar ratio, but it must be both small enough and non-zero for the model consistency.  From Eq. (24), we can directly read several key points of $\eta$ and $\eta'$, at the time of the inflation finish there is $\eta(0)=\frac{3}{2}=\epsilon(0)$, this is also a requirement of the model consistency.

 Now we start from Eq. (24), then we can easily solve the inflationary evolutions. Firstly, from the $\eta$ definition formula in Eq. (17), we can solve out $\rho_{\Phi_{DM}}(N)$ by integrating $\eta(N)$. Secondly, we can further integrate $\rho_{\Phi_{DM}}(N)$ to obtain $\rho_{\Phi}(N)$ by use of the first formula in Eq. (17). Thirdly, $\epsilon(N)$ is given by the second formula in Eq. (15). The derived results are
\ba
 \frac{\rho_{\Phi_{DM}}(N)}{\rho_{\Phi}(0)}=e^{2\int^{N}_{0}\eta(N')dN'},\hspace{0.3cm}
 \frac{\rho_{\Phi}(N)}{\rho_{\Phi}(0)}=1+3\int^{N}_{0}\frac{\rho_{\Phi_{DM}}(N')}{\rho_{\Phi}(0)}dN',\hspace{0.3cm}
 \epsilon(N)=\frac{3}{2}\,\frac{\rho_{\Phi_{DM}}(N)}{\rho_{\Phi}(N)}\,,
\ea
 where there is $\rho_{\Phi_{DM}}(0)=\rho_{\Phi}(0)=3\tilde{M}_{p}^{2}H_{inf}^{2}$ and $\epsilon(0)=\frac{3}{2}$ at the end of the inflation. Lastly, we can further calculate $\rho_{\Phi_{DE}}=\rho_{\Phi}-\rho_{\Phi_{DM}}$, $T_{\Phi}=|\dot{\Phi}|^{2}=\frac{1}{2}\rho_{\Phi_{DM}}$, $V_{\Phi}=\rho_{\Phi}-T_{\Phi}$, and also $\eta_{V}$, $w_{\Phi}$, etc., note that all kinds of the energy forms are now normalized to $\rho_{\Phi}(0)$.

 Now we show the numerical results of this inflation model. Fig. 4 shows the inflationary evolutions of the relevant energy forms of the $\Phi$ field with $N$ as time scale.
\begin{figure}
 \centering
 \includegraphics[totalheight=7cm]{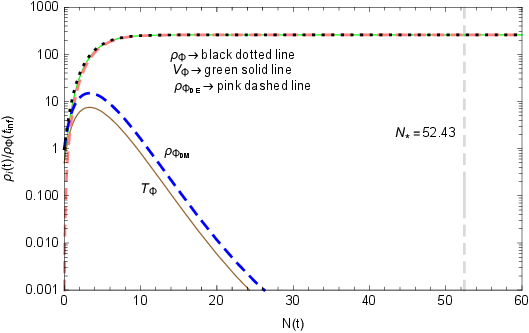}
 \caption{The inflationary evolutions of the relevant energy forms of the $\Phi$ field with the e-fold number as time scale, $N_{*}\approx52.43$ is corresponding to the time of $k_{*}=0.05$ $\mathrm{Mpc}^{-1}$ exiting from horizon.}
\end{figure}
 In the early and middle phases of the inflation process, these three curves of $\rho_{\Phi},\rho_{\Phi_{DE}},V_{\Phi}$ almost coincide with each other, moreover, they nearly keep a constant value. The reason for this is that this stage is the $\Phi_{DE}$-dominated phase, the $\Phi_{DM}$ and $T_{\Phi}$ growths are very slow, they are only yielded to very small amounts, so there is $\dot{\Phi}\approx0$, this is namely so-called slow-roll inflation. In the later phase of the inflation proceeding, the $\Phi_{DM}$ and $T_{\Phi}$ growths are more and more fast and reach their peak values, thereby $\rho_{\Phi}$, $\rho_{\Phi_{DE}}$ and $V_{\Phi}$ turn into fall sharply and their curves are significantly separated each other. Eventually, $\rho_{\Phi_{DM}}$ exceeds $\rho_{\Phi_{DE}}$, the $\Phi_{DE}$-dominated universe is thus transformed into the $\Phi_{DM}$-dominated one, at the same time, the accelerating expansion is also changed into the decelerating one, as a result, the inflation is naturally and smoothly terminated. At the time of the inflation beginning, namely at $N=+\infty$, there are $\rho_{\Phi_{DE}}=\rho_{\Phi}=V_{\Phi}\approx258.9\rho_{\Phi}(0)$ and $\rho_{\Phi_{DM}}=T_{\Phi}=0$. At the time of the inflation finish, namely at $N=0$, there are $\rho_{\Phi_{DE}}=0$ and $\rho_{\Phi_{DM}}=\rho_{\Phi}=2T_{\Phi}=2V_{\Phi}$. In short, Fig. 4 clearly shows the full evolutions of all kinds of the energy forms in the inflation process, from which we can see the slow-roll features, the transformations among these energy forms, and the inflation terminations.

\begin{figure}
 \centering
 \includegraphics[totalheight=7cm]{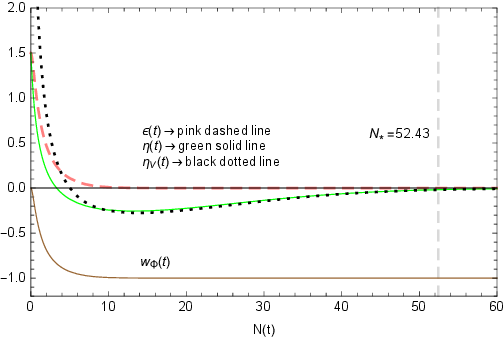}
 \caption{The inflationary evolutions of the slow-roll parameters and parameter-of-state with the e-fold number as time scale, $N_{*}\approx52.43$ is corresponding to the time of $k_{*}=0.05$ $\mathrm{Mpc}^{-1}$ exiting from horizon.}
\end{figure}
 Fig. 5 shows the inflationary evolutions of the slow-roll parameters and parameter-of-state with $N$ as time scale. We can see three remarkable features. i) In the most of the inflation duration, there are $\epsilon\ll1$ and $w_{\Phi}\approx-1$, this is exactly the $\Phi_{DE}$-dominated phase. When N is approaching to $N(t_{inf})=0$, $\epsilon$ and $w_{\Phi}$ sharply rise, the inflation enters the $\Phi_{DM}$-dominated phase, and it is soon terminated as $\Phi_{DE}$ is exhausted. ii) In the $\Phi_{DE}$-dominated phase $\eta$ is almost coinciding with $\eta_{V}$ due to $\eta\approx\eta_{V}$ (see Eq. (21)), but they are distinctly separated each other in the $\Phi_{DM}$-dominated phase. iii) $\epsilon$ is always positive, whereas $\eta$ and $\eta_{V}$ can change from negative to positive as the inflation is close to its end. $M^{2}_{\Phi}$ is proportional to $\eta_{V}$ in Eq. (23), this means that $M_{\Phi}$ is gradually generated from nothing as $\eta_{V}$ evolving from negative to positive, undoubtedly, this is owing to $\Phi_{DM}$ growing to a certain amount, therefore $M_{\Phi}$ purely results from the inflationary dynamical evolution. This mechanism of the inflationary field mass generation is very different from that of the SM particle mass, which simply arises from the vacuum spontaneous breaking. In a word, the whole physical mechanism and picture of the primordial inflation are excellently explained by Fig. 4 and Fig. 5.

 Next we set about addressing the observed data of the inflation. In Fig. 5, at $N_{*}\approx52.43$ the slow-roll parameters and the parameter-of-state are evaluated as follows,
\ba
 \epsilon_{V*}\approx\epsilon_{*}\approx1.05\times10^{-7},\hspace{0.3cm} \eta_{*}\approx-0.0175,\hspace{0.3cm} \eta'_{*}\approx0.0020,\hspace{0.3cm} \eta_{V*}\approx-0.0183,\hspace{0.3cm} w_{\Phi*}\approx-1,
\ea
 where hereinafter the ``$*$" subscript specially indicates the time of $N_{*}\approx52.43$. From the cosmological perturbation theory of the structure formation \cite{26}, we know that the above slow-roll parameters are directly related to the following inflationary observable quantities,
\begin{alignat}{1}
 &\Delta^{2}_{R}(k_{*})=\frac{H^{2}}{8\pi^{2}\tilde{M}^{2}_{p}\,\epsilon}|_{k_{*}}
   =\frac{1}{8\pi^{2}\epsilon_{*}}[\frac{\rho_{\Phi}(N_{*})}{\rho_{\Phi}(0)}][\frac{H_{inf}}{\tilde{M}_{p}}]^{2},\hspace{0.5cm}
   r_{*}=16\epsilon_{*}\,,\nonumber\\
 &n_{s}(k_{*})-1=\frac{dln\Delta^{2}_{R}}{dlnk}|_{k_{*}}=\frac{dlnH^{2}-dln\epsilon}{(\epsilon-1)dN}|_{k_{*}}
                         =-4\epsilon_{*}+2\eta_{*}\approx6(\frac{\eta'_{*}}{9}-\epsilon_{V*})+2\eta_{V*}\,,\nonumber\\
 &\frac{dn_{s}}{dlnk}|_{k_{*}}=\frac{dn_{s}}{(\epsilon-1)dN}|_{k_{*}}=4\epsilon'_{*}-2\eta'_{*}
                                       \approx24\epsilon_{V*}(\frac{2\eta'_{*}}{9}-\epsilon_{V*})+16\epsilon_{V*}\eta_{V*}-2\xi^{2}_{V*}\,,
\end{alignat}
 where I employ Eq. (21). $k_{*}=0.05$ $\mathrm{Mpc}^{-1}$ is the pivot scale exciting from horizon at $N_{*}\approx52.43$, the relation between $k_{*}$ and $N_{*}$ will be given by the following Eq. (28). $H_{inf}$ is the expansion rate at the time of the inflation finish, it is also an input parameter in the inflation sector, we can determine $H_{inf}\approx1.99\times10^{10}$ GeV by fitting the scalar power spectra. $\frac{\rho_{\Phi}(N_{*})}{\rho_{\Phi}(0)}\approx258.9$ is calculated out by Eq. (25), also see Fig. 4. Note that the $\frac{\eta'_{*}}{9}$ term in Eq. (27) is the same order of magnitude as $\epsilon_{V*}$, but the both contributions are actually negligible. We can employ either of the two sets of slow-roll parameters to calculate the inflationary data, put Eq. (26) into Eq. (27), we can precisely reproduce all of the measured inflation data in Eq. (1), finally, we can also calculate out $M_{\Phi}\approx5.05\times10^{10}$ GeV by Eq. (23). All kinds of the input parameters and the output results are all summarized in Table 2 in VI Section.

 $k_{*}$ is defined and calculated as follows,
\ba
 k_{*}=\frac{a_{*}H_{*}}{c}=[\frac{H_{0}}{c\,h}][\frac{g_{*}(T_{0})}{2}]^{\frac{1}{3}}[\Omega_{\gamma}(T_{0})h^{2}]^{\frac{1}{3}}
 [\frac{H_{inf}}{H_{0}/h}]^{\frac{1}{3}}[\frac{T_{re}}{T_{0}}]^{\frac{1}{3}}[\frac{\rho_{\Phi}(N_{*})}{\rho_{\Phi}(0)}]^{\frac{1}{2}}e^{-N_{*}},
\ea
 where $c$ is the speed of light, $\frac{H_{0}}{ch}=\frac{100}{3\times10^{5}}$ $\mathrm{Mpc}^{-1}$ and $H_{0}/h\approx2.13\times10^{-42}$ GeV are two fixed constants, a detailed derivation of Eq. (28) is seen in Appendix I. At the present day, the CMB temperature is $T_{0}\approx2.7255\;\mathrm{K}\approx2.35\times10^{-4}$ eV \cite{1}, the effective number of relativistic degrees of freedom is $g_{*}(T_{0})\approx4.1$ (which includes the $\nu_{R}$ contribution, see the following Eq. (38)), the photon energy density parameter is $\Omega_{\gamma}(T_{0})h^{2}\approx2.47\times10^{-5}$, which is obtained by Eq. (49) in Section V, in addition, we will calculate out the reheating temperature $T_{re}\approx8.29\times10^{10}$ GeV by Eq. (36) in Section IV. After the above quantities are input into Eq. (28), we can immediately work out $N_{*}\approx52.43$ corresponding to $k_{*}=0.05$ $\mathrm{Mpc}^{-1}$. It should be emphasized that Eq. (28) relates these fundamental quantities of the inflation, reheating and current universe together, this is rightly another characteristic of the unified model.

 Finally, we draw a profile of the inflationary potential $V(|\Phi|/\tilde{M}_{p})$ and find out its analytical function form. Since the $\epsilon(N)$ solution has been obtained, see Fig. 5, substitute it into Eq. (20) and make a numerical integration, then we can obtain a numerical solution of $|\Phi(N)|/\tilde{M}_{p}$, on the other hand, the $V_{\Phi}(N)$ solution has been given in Fig. 4, put these two solutions together, then we can translate $V_{\Phi}(N)$ with $N$ as variable into $V(|\Phi|/\tilde{M}_{p})$ with $|\Phi|/\tilde{M}_{p}$ as variable, this is easily accomplished by a computer, the calculated result is shown by the green solid curve in Fig. 6.
\begin{figure}
 \centering
 \includegraphics[totalheight=7cm]{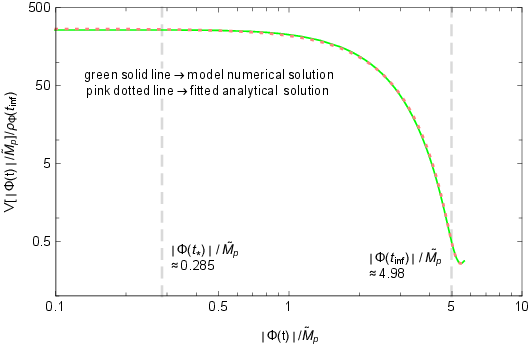}
 \caption{The inflationary potential evolution with $\frac{|\Phi(t)|}{\tilde{M}_{p}}$ as variable. $t_{*}$ and $t_{inf}$ are respectively corresponding to $N_{*}\approx52.43$ and $N=0$. The left end of the curves is the $\Phi_{DE}$-dominated phase, while the right end of the curves is the $\Phi_{DM}$-dominated phase, the small concavity is the minimum of the potential energy, in which $\Phi_{DM}$ takes oscillation and decay. The green solid curve is the numerical solution of the model, while the pink dotted curve is the analytical solution of Eq. (29), the latter is perfectly fitting the former.}
\end{figure}
 $|\Phi(t_{*})|/\tilde{M}_{p}\approx0.285$ and $|\Phi(t_{inf})|/\tilde{M}_{p}\approx4.98$ are two field values corresponding to $N_{*}\approx52.43$ and $N=0$, respectively. When $t\rightarrow0$ (namely $N\rightarrow+\infty$), there is $|\Phi(t)|/\tilde{M}_{p}\rightarrow0$ and $V(|\Phi(t)|/\tilde{M}_{p})/\rho_{\Phi}(t_{inf})\rightarrow258.9$, this is the $\Phi_{DE}$-dominated phase. When $t\rightarrow t_{inf}$ (namely $N\rightarrow0$), there is $|\Phi(t)|/\tilde{M}_{p}\rightarrow4.98$ and $V(|\Phi(t)|/\tilde{M}_{p})/\rho_{\Phi}(t_{inf})\rightarrow0.5$, this is the $\Phi_{DM}$-dominated phase. When $t>t_{inf}$\,, namely $-1\leqslant N<0$, there is a small concavity of the potential energy, at $|\Phi(t)|/\tilde{M}_{p}\approx5.46$ the potential energy has the minimum value of $V(|\Phi(t)|/\tilde{M}_{p})/\rho_{\Phi}(t_{inf})\approx0.261$. At $|\Phi(t)|/\tilde{M}_{p}\approx5.64$ there is $\dot{\Phi}=0$, namely the kinetic energy vanishing, thereby the $\Phi$ field reaches its maximum value, after then it will turn back, thus it can oscillate back and forth in the small concavity, and eventually decay into the other particles.

 By means of fitting the numerical solution of $V(|\Phi(t)|/\tilde{M}_{p})$, I find its analytical function form such as
\begin{alignat}{1}
 &\frac{V(x^{2})}{\rho_{\Phi}(t_{inf})}=0.2612\left[1+a(x^{2}-x_{e}^{2})+b(x^{2}-x_{e}^{2})^{2}\right]e^{-a(x^{2}-x_{e}^{2})},\\
 &x=\frac{|\Phi(t)|}{\tilde{M}_{p}}\,,\hspace{0.5cm} x_{e}\approx5.462,\hspace{0.5cm} a\approx0.132,\hspace{0.5cm}  b\approx0.02575,\nonumber\\
 \Longrightarrow\; &\frac{dV}{dx}|_{x_{e}}=0,\hspace{0.3cm} \frac{V_{min}}{\rho_{\Phi}(t_{inf})}|_{x_{e}}=0.2612,\hspace{0.3cm}  
  \frac{V}{\rho_{\Phi}(t_{inf})}|_{x=4.981}=0.5,\hspace{0.3cm} \frac{V_{max}}{\rho_{\Phi}(t_{inf})}|_{x=5.644}=0.28,\nonumber
\end{alignat}
 where $\rho_{\Phi}(t_{inf})=3\tilde{M}_{p}^{2}H_{inf}^{2}$ as before, $a$ and $b$ are two adjustable parameters, the other values are all fixed. The potential form of Eq. (29) consists of two factor of the usual scalar potential and the Gauss-type exponential potential, so it explicitly satisfies the model requirement in Eq. (2), note that it is very different from a wide variety of the inflationary potentials discussed by the early papers, for example, the Starobinsky-type potential \cite{27}. In Fig. 6, we use the pink dotted curve to show the analytical solution of Eq. (29), it can perfectly fit the model numerical solution shown by the green solid curve. When $t\rightarrow0$,  there is $x\rightarrow0$ and $\frac{V}{\rho_{\Phi}(t_{inf})}\rightarrow258.9$, this is purely the primordial dark energy, which is really the source of the universe evolution, all things including the $\Phi$ field, the field motion $\dot{\Phi}$ and the primordial $\Phi_{DM}$ grow from it.
 
 If we now start from Eq. (29) to deal with the inflation problem by use of the traditional procedure, in principle, we can also derive all kinds of the results which are early obtained by my ansatz, but the potential form of Eq. (29) can not at all be guessed in advance. In conclusion, Fig. 6 clearly shows the inflationary evolution and Eq. (29) directly gives us deep insights into the inflationary potential characterization, in particular, the physical mechanism and essence of the inflation are very well understood in the unified model. Up to now, all of the inflation problems have completely been solved.

\vspace{0.6cm}
\noindent\textbf{IV. Reheating and Baryogenesis}

\vspace{0.3cm}
 At the end of the inflation, $\Phi_{DE}$ is completely exhausted, the universe is entirely filled by $\Phi_{DM}$ and it turns into the decelerating expansion era. Because $\Phi_{DM}$ is an excited particle state with the super-heavy mass, it has kinetic energy and no pressure, it can interact with the other particles through those couplings in Eq. (2), therefore $\Phi_{DM}$ is quite unstable and can shortly decay into one SM particle and one dark particle. The phenomena of the $\Phi_{DM}$ decay has been discussed in Section II, in this Section we will describe the important cosmological implications of the $\Phi_{DM}$ decay, namely, it directly brings about the universe reheating to produce the hot bath, simultaneously it also generates the matter-antimatter asymmetry through the aforementioned leptogenesis mechanism.

 From now on, we take off the ``$DM$" subscript of $\Phi_{DM}$, directly use $\Phi$ to indicate $\Phi_{DM}$ since $\Phi_{DE}$ has vanished. The $\Phi$ decay and its subsequent processes fully produce the earliest radiation of the universe, which are a hot plasma consisting of the SM particles and the dark ones. The total energy of the universe now includes the two components of $\rho_{\Phi}$ and $\rho_{R}$. The dynamic of the reheating evolution are collectively controlled by Friedmann equation and the continuity equations, namely
\ba
 \rho_{\Phi}+\rho_{R}=3\tilde{M}^{2}_{p}H^{2},\hspace{0.5cm} \dot{\rho}_{\Phi}+3H\rho_{\Phi}=-\Gamma_{\Phi}\rho_{\Phi},\hspace{0.5cm}
 \dot{\rho}_{R}+4H\rho_{R}=\Gamma_{\Phi}\rho_{\Phi},
\ea
 where $\Gamma_{\Phi}$ is the $\Phi$ decay width, it has been calculated by the particle model in Eq. (7), I take $\frac{\mu_{0}}{M_{\Phi}}\approx0.005$ as a suitable input value, which is the only input parameter in the reheating sector, see the following Table 2. The physical implications of Eq. (30) are very clear, it is a closed system of equations once $\Gamma_{\Phi}$ is provided, the evolutions of $H$, $\rho_{\Phi}$ and $\rho_{R}$ are completely determined by the $\Gamma_{\Phi}$ value and the initial conditions.

 In order to solve the system of equations of Eq. (30), we can introduce the dimensionless energy densities and time variable as follows,
\begin{alignat}{1}
 &\tilde{\rho}_{i}(\tilde{t})=\frac{\rho_{i}}{3\tilde{M}^{2}_{p}\Gamma_{\Phi}^{2}}\,,\hspace{0.3cm}
   \tilde{t}=(t-t_{inf})\Gamma_{\Phi}=\frac{t-t_{inf}}{\tau_{\Phi}}\,,\hspace{0.3cm} 0<\tilde{t}\leqslant\tilde{t}_{ref}=\frac{t_{ref}-t_{inf}}{\tau_{\Phi}},\\
 &\Longrightarrow \tilde{\rho}_{\Phi}(0)=(\frac{H_{inf}}{\Gamma_{\Phi}})^{2},\hspace{0.5cm} \tilde{\rho}_{R}(0)=0\,,
\end{alignat}
 where $i=(\Phi,R$) and $\tau_{\Phi}$ is the $\Phi$ lifetime. The time of the reheating beginning is namely the time of the inflation finish, while the time of the reheating finish is specially indicated by the ``ref" subscript. Eq. (32) is exactly the initial values of the reheating evolution. The numerical values in Table 2 gives $\frac{H_{inf}}{\Gamma_{\Phi}}\approx7.9\times10^{5}$, this means that the reheating is severely out-of-equilibrium in its early phase. By use of Eq. (31), we can recast Eq. (30) as follows,
\ba
 \tilde{\rho}_{\Phi}+\tilde{\rho}_{R}=(\frac{H}{\Gamma_{\Phi}})^{2},\hspace{0.3cm}
 \frac{d\tilde{\rho}_{\Phi}}{d\tilde{t}}+[3(\frac{H}{\Gamma_{\Phi}})+1]\tilde{\rho}_{\Phi}=0\,,\hspace{0.3cm}
 \frac{d\tilde{\rho}_{R}}{d\tilde{t}}+4(\frac{H}{\Gamma_{\Phi}})\tilde{\rho}_{R}-\tilde{\rho}_{\Phi}=0\,.
\ea
 Now the evolutions of$\frac{H}{\Gamma_{\Phi}}$, $\tilde{\rho}_{\Phi}(\tilde{t})$ and $\tilde{\rho}_{R}(\tilde{t})$ only depend on the initial values in Eq. (32), therefore we can easily obtain their numerical solutions.

 In fact, we are much more interested in the evolutions of the energy density parameters and total parameter-of-state, which are defined by
\ba
 \Omega_{i}(\tilde{t})=\frac{\rho_{i}}{\rho_{\Phi}+\rho_{R}}\,,\hspace{0.5cm}
 w_{T}(\tilde{t})=\frac{P_{\Phi}+P_{R}}{\rho_{\Phi}+\rho_{R}}=\frac{\Omega_{R}}{3}\,,
\ea
 where $P_{\Phi}=0$ and $P_{R}=\frac{\rho_{R}}{3}$.
\begin{figure}
 \centering
 \includegraphics[totalheight=7cm]{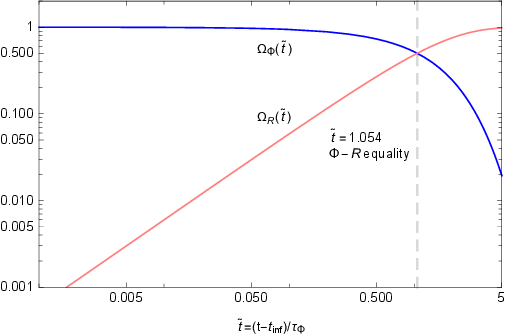}
 \caption{The reheating evolutions of the energy density parameters of the $\Phi$ and radiation with the dimensionless $\tilde{t}$ as time scale. $\tilde{t}\approx1.054$ is the time of $\rho_{\Phi}=\rho_{R}$ and $\tilde{t}\approx5$ is nearly the time of the reheating finish.}
\end{figure}
 Fig. 7 numerically shows the reheating evolutions of $\Omega_{\Phi}$ and $\Omega_{R}$ with $\tilde{t}$ as time scale. As more and more $\Phi$ decay, the $\Phi$ density parameter is continuously decreasing from the initial $\Omega_{\Phi}=1$ to the final $\Omega_{\Phi}\rightarrow0$, $\Phi$ is eventually exhausted, meanwhile, $\Omega_{R}$ is gradually increasing from the initial $\Omega_{R}=0$ to the final $\Omega_{R}\rightarrow1$, at the end of the reheating the universe is entirely filled by radiation. The $w_{T}(\tilde{t})$ evolution is also in agreement with this, which is gradually increasing from the initial $w_{T}=0$ to the final $w_{T}=\frac{1}{3}$\,, as a result, the $\Phi$-dominated universe is gradually transformed into the R-dominated one. In the reheating evolution, the expansion rate is continuously declining, and the universe is decelerating expansion.

 In Fig. 7, the time of the $\Phi-R$ equality is $\tilde{t}_{req}\approx1.054$, it is a key time point in the reheating process. When $0<\tilde{t}\leqslant\tilde{t}_{req}$ (namely $t_{inf}<t\leqslant t_{req}$), the universe is the $\Phi$-dominated phase, once $\tilde{t}>\tilde{t}_{req}$ ($t>t_{req}$) the universe turns into the radiation-dominated phase, at $\tilde{t}\approx5$ the reheating process is nearly finished, after then the universe will start a new era, namely the hot expansion driven by the radiation. On the other hand, $\tilde{t}_{req}\approx1.054$ means $t_{req}-t_{inf}\approx\tau_{\Phi}$, this indicates that the $\Phi$ decay mostly happens around $t_{req}$\,, $\rho_{R}(t_{req})$ is essentially the highest value of the radiation energy density, thereby the corresponding temperature $T_{re}(t_{req})$ is the highest temperature in the reheating process. To sum up, $\tilde{t}_{req}$ and $T_{re}$ are determined by the following relations,
\begin{alignat}{1}
 &\Omega_{\Phi}(\tilde{t}_{req})=\Omega_{R}(\tilde{t}_{req})=\frac{1}{2}\;\Longrightarrow\tilde{t}_{req}
   =\frac{t_{req}-t_{inf}}{\tau_{\Phi}}\approx1.054\,,\\
 &\frac{H(t_{req})}{\Gamma_{\Phi}}=\sqrt{2\tilde{\rho}_{R}(\tilde{t}_{req})}\,,\hspace{0.3cm}
   T_{re}=[\tilde{M}_{p}H(t_{req})]^{\frac{1}{2}}\left[\frac{45}{\pi^{2}g_{*}(T_{re})}\right]^{\frac{1}{4}}
   =[\tilde{M}_{p}\Gamma_{\Phi}]^{\frac{1}{2}}\left[\frac{90\tilde{\rho}_{R}(\tilde{t}_{req})}{\pi^{2}g_{*}(T_{re})}\right]^{\frac{1}{4}},
\end{alignat}
 where $g_{*}(T_{re})=121$ is the effective number of relativistic degrees of freedom, which includes all of the model particles except $\Phi$. The numerical calculation gives $\frac{H(t_{req})}{\Gamma_{\Phi}}\approx0.58$ and $T_{re}\approx8.29\times10^{10}$ GeV, this demonstrates that the reheating is really out-of-equilibrium in the period of $t_{inf}<t<t_{req}$\,, after $t>t_{req}$ the thermal equilibrium is shortly formed. At $\tilde{t}=5$ there is $T(t_{ref})\approx4.22\times10^{10}\,\mathrm{GeV}<M_{\Phi}\approx5.05\times10^{10}\,\mathrm{GeV}$, the $\Phi$ particle can not be produced from the hot bath anew, thus the reheating is naturally over. The relevant results of the reheating are all listed in Table 2 in Section VI. Finally, we stress that the reheating is closely related to the inflation and the particle physics, this is an important characteristic of the unified model.

 Now we discuss the baryogenesis through the foregoing leptogenesis mechanism. On the basis of the discussions in Section II, the decay $\Phi\rightarrow l^{c}+\nu_{R}$ is a relatively weak decay mode of $\Phi$, but it has the three remarkable features. i) Its decay rate has the $CP$ asymmetry about $A_{CP}\sim10^{-10}$, which is obtained by Eq. (7). ii) This decay is indeed out-of-equilibrium in the reheating process since $\Gamma(\Phi\rightarrow l^{c}+\nu_{R})\ll\Gamma_{\Phi}< H(t)$ $(t_{inf}<t\leqslant t_{req})$. iii) Although the net lepton number is conserved as zero, the $CP$ asymmetric decay can generate the anti-lepton asymmetry and the $\nu_{R}$ asymmetry with the same amount. After the reheating is completed, these two asymmetries are respectively isolated in the SM sector and dark sector, so they can not be erased each other. In the SM sector the anti-lepton asymmetry can partly be converted into the baryon asymmetry through the sphaleron transition \cite{24}, whereas in the dark sector the $\nu_{R}$ asymmetry is forever frozen. In short, although this baryogenesis mechanism does not fully fulfil the Sakharov's three conditions \cite{28}, it is indeed put into effect in the unified model.

 After the hot bath has fully been formed, the universe enters into the hot expansion era. The hot evolution in the dark sector will be discussed in Section V, while the hot evolution in the SM sector is exactly the well-known hot big bang paradigm. Because the $B-L$ number is always conserved, above the electroweak scale the sphaleron transition with $B+L$ violation can partly convert the anti-lepton asymmetry into the baryon asymmetry \cite{29}. The detailed relations are given as follows,
\begin{alignat}{1}
 &[\frac{\bar{n}_{l}-n_{l}}{s}]_{T_{re}}=[\frac{n_{\nu_{R}}-\bar{n}_{\nu_{R}}}{s}]_{T_{re}}=[\frac{n_{\Phi}A_{CP}}{s}]_{T_{re}}
   =[\frac{\rho_{\Phi}A_{CP}}{M_{\Phi}s}]_{T_{re}}=[\frac{\rho_{R}}{s}]_{T_{re}}\frac{A_{CP}}{M_{\Phi}}=\frac{3T_{re}A_{CP}}{4M_{\Phi}}\,,\nonumber\\
 &\eta_{B}=[\frac{s}{n_{\gamma}}]_{T_{0}}Y_{B}(T_{0})=[\frac{s}{n_{\gamma}}]_{T_{0}}Y_{B}(T_{ew})
   =[\frac{s}{n_{\gamma}}]_{T_{0}}c_{s}Y_{B-L}(T_{ew})=[\frac{s}{n_{\gamma}}]_{T_{0}}c_{s}Y_{B-L}(T_{re})\nonumber\\
 &\hspace{0.5cm}=c_{s}[\frac{s}{n_{\gamma}}]_{T_{0}}[\frac{\bar{n}_{l}-n_{l}}{s}]_{T_{re}},
\end{alignat}
 where $s$ is the entropy density and $Y_{B}(T)=[\frac{n_{B}-\bar{n}_{B}}{s}]_{T}$ is the yield, $c_{s}=\frac{28}{79}$ is the SM sphaleron coefficient, $[\frac{s}{n_{\gamma}}]_{T_{0}}\approx7.38$ since $g_{*}(T_{0})\approx4.1$ includes the $\nu_{R}$ contribution (see Eq. (38)). Eq. (37) clearly shows that $\eta_{B}$ is collectively determined by the inflaton mass, the reheating temperature and the $CP$ asymmetry, this again manifests that the unified model closely relates the inflation, reheating and particle physics together. In Table 2, I take $A_{CP}\approx1.91\times10^{-10}$ as an input value provided by the particle model, then the unified model naturally predicts $\eta_{B}\approx6.14\times10^{-10}$.

\vspace{0.6cm}
\noindent\textbf{V. Current Dark Matter and Dark Energy}

\vspace{0.3cm}
 The hot expansion of the universe causes that the radiation energy is red-shifted, therefore the universe temperature, which is indicated by the photon temperature, is continuously declining. The hot evolution in the dark sector is however different from that in the SM sector. On the basis of the discussion in Section II, these heavy dark particles of $N^{0},E^{-},\phi^{0}$ are all depleted by their decays in the very early universe, only the stable $S$ and $\nu_{R}$ can survive in the dark sector. In fact, $\nu_{R}$ has been decoupled from the hot bath below the energy scale of $M_{N}\sim10^{8}$ GeV, while the $S$ annihilation is frozen out below the temperature of $T_{f}\approx\frac{M_{S}}{25.4}\approx10.1$ GeV, thus the dark sector is eventually separated from the SM sector. As a result, the relativistic $\nu_{R}$ becomes the dark radiation background, while the non-relativistic $S$ becomes the CDM. 
 
 From the entropy conservation, we can derive the effective temperature of $\nu_{R}$ as follows,
\begin{alignat}{1}
 &\frac{a^{3}(M_{N})}{a^{3}(T)}=\frac{T_{\nu_{R}}^{3}}{M_{N}^{3}}=\frac{g_{*}(T)T^{3}}{g_{*}(M_{N})M_{N}^{3}}\,,\nonumber\\
 \Longrightarrow\; &(\frac{T_{\nu_{R}}}{T})^{3}=\frac{2+\frac{7}{8}\times6[(\frac{T_{\nu_{L}}}{T})^{3}+(\frac{T_{\nu_{R}}}{T})^{3}]}{117.5}\;
 \Longrightarrow (\frac{T_{\nu_{R}}}{T})^{3}\approx0.0348\,,
\end{alignat}
 where $(\frac{T_{\nu_{L}}}{T})^{3}=\frac{4}{11}$ is the well-known effective temperature of $\nu_{L}$ in the SM sector, and there must be $T<m_{e}\approx0.5$ MeV (namely after the electron-positron annihilation). From Eq. (38), we can obtain the present-day $T_{\nu_{R}}=0.0348^{\frac{1}{3}}T_{0}\approx0.9$ K, moreover, we can calculate the effective number of neutrinos at the recombination as $N_{eff}=3[1+(T_{\nu_{R}}/T_{\nu_{L}})^{4}]\approx3.13$, which is safely within the current limit given by the CMB data analysis.

 However, the CDM $S$ has a special nature similar to the inflation field. On account of the momentum red-shift, the effective temperature of $S$ (namely its kinetic temperature) scales as $T_{S}\propto E_{S}\propto p_{S}^{2}\propto a^{-2}$ where $E_{S}$ and $p_{S}$ are respectively the kinetic energy and momentum of $S$, by contrast $T_{\nu_{R}}\propto a^{-1}$, therefore $T_{S}$ is much faster falling than $T_{\nu_{R}}$\,. At the present day, $T_{S}$ is almost approaching to absolute zero, namely, the $S$ kinetic energy is essentially exhausted, so $S$ actually becomes a supercool matter, thus the supercool $S_{DM}$ can gradually condense into the dark energy $S_{DE}$ via its special self-interacting potential. This phenomenon is exactly a cosmological effect of the Bose–Einstein condensate, which occurs at the extremely low temperature in general.

 When the universe temperature drops to $T_{eq}\approx1$ eV, namely the universe age is about $5\times10^{4}$ years, the total matter density exceeds the radiation one, thus the universe is transformed from the radiation-dominated era to the matter-dominated era. Note that $T_{eq}$ is below $T_{BBN}\approx0.1$ MeV but above $T_{Recom}\approx0.3$ eV. We can regard the time of the matter-radiation equality as the starting point of $S_{DM}$ condensing into $S_{DE}$. As the universe is more and more cooling, the $S_{DM}$ effective temperature is rapidly dropping and approaching to absolute zero, accordingly the $S_{DM}$ kinetic energy is exhausted fast, thus more and more $S_{DM}$ become the supercool matter so that they eventually condense into $S_{DE}$, in other words,  more and more $S_{DE}$ are growing from them. On the physical picture, the $S$ condensation is essentially a reverse process of the $\Phi$ inflation discussed in III Section, namely it is actually a process of $S_{DE}$ slowly growing from $S_{DM}$, or $S_{DM}$ gradually converting into $S_{DE}$. It should be stressed that the $S$ condensation is very different from baryon and electron condensing into the usual material in the visible world, the former is a pure boson system, whereas the latter is a pure fermion system. In a word, it is the special evolution of $S$ to lead to the dark matter and dark energy in the current universe.

 From now on, we directly use the abbreviations of ``$DM$" and ``$DE$" to denote $S_{DM}$ and $S_{DE}$, respectively, their energy density and pressure are given as follows,
\begin{alignat}{1}
 &P_{DM}=0\,,\hspace{0.3cm} P_{DE}=-\rho_{DE}\,,\hspace{0.3cm} \rho_{DM}+\rho_{DE}=\rho_{S}\,,\hspace{0.3cm}
   P_{DM}+P_{DE}=P_{S}=w_{S}\rho_{S}\,,\nonumber\\
 &\Longrightarrow \rho_{DM}=(1+w_{S})\rho_{S}\,,\hspace{0.5cm} \rho_{DE}=-w_{S}\rho_{S}\,,
\end{alignat}
 where $w_{S}$ is a parameter-of-state varying with the time, and there is $0\geqslant w_{S}\geqslant-1$. The physical implications of Eq. (39) is parallel to that of Eq. (10), see those explanations for the $\Phi$ field below Eq. (10). After $T<T_{eq}\approx1$ eV, the universe energy includes the four components of the photon $\rho_{\gamma}$, the neutrino $\rho_{\nu}$ (which includes the $\nu_{L}$ and $\nu_{R}$ energy), the baryon $\rho_{B}$, and the dark $\rho_{S}$ (which consists of $\rho_{DM}$ and $\rho_{DE}$), their dynamical evolutions are determined by the following system of equations,
\begin{alignat}{1}
 &\rho_{\gamma}+\rho_{\nu}+\rho_{B}+\rho_{S}=3\tilde{M}^{2}_{p}H^{2},\\
 &\dot{\rho}_{\gamma}+4H\rho_{\gamma}=0\,,\hspace{0.3cm} \dot{\rho}_{\nu}+3H\rho_{\nu}(1+w_{\nu})=0\,,\hspace{0.3cm}
  \dot{\rho}_{B}+3H\rho_{B}=0\,,\nonumber\\
 &\dot{\rho}_{S}+3H\rho_{S}(1+w_{S})=0\:\Longrightarrow\dot{\rho}_{DE}=-(\dot{\rho}_{DM}+3H\rho_{DM}),\\
 &\dot{\rho}_{DE}=\kappa(T)H\rho_{DM}\,.
\end{alignat}
 Eq. (40) is Friedmann equation, it relates the SM sector and the dark sector together, and it controls the expansion rate. The four equations in Eqs. (41) are respectively the continuity equations of the four energy components. The neutrino is relativistic state in the early phase, but it can turn into non-relativistic state in the later phase because of its a sub-eV mass, so there is $w_{\nu}=\frac{1}{3}$ for the relativistic neutrino and $w_{\nu}=0$ for the non-relativistic neutrino. The last equality in Eqs. (41) indicates that the $S_{DE}$ growth is purely from the $S_{DM}$ reduction in the comoving volume. Eq. (42) is namely the growth equation of $S_{DE}$, the parameter $\kappa(T)$ characterizes the growth rate (or one may also call it as the condensing rate), it explicitly depends on the temperature, $\kappa$ will rapidly increase as $T$ is more and more low. Once the evolution function of $\kappa(T)$ is provided, then the above system of equations is completely closed, thus we can solve the evolution of each energy component.

 The $S$ condensation is also a very slow process similar to the inflation evolution, therefore we can also introduce the e-fold number of the condensation process as the time scale, which is defined by
\begin{alignat}{1}
 &N(T)=ln\frac{a(T)}{a(T_{eq})}=ln\frac{T_{eq}}{T}\:\Longrightarrow \dot{N}(t)=H(t),\\
 &a(T_{eq})\leqslant a(T)\leqslant a(T_{0})=1,\hspace{0.3cm} 0=N(T_{eq})\leqslant N(T)\leqslant N(T_{0})=N_{0},\nonumber
\end{alignat}
 where $T_{eq}$ is the starting temperature of the $S$ condensation and $T_{0}$ is the present-day universe temperature. Note that $\dot{N}(t)$ is positive in Eq. (43), namely $N$ is increasing with the time, one should not confuse it with the negative $\dot{N}(t)$ defined in Eq. (16).

 Now we use $N$ as the time variable of the energy evolution, and normalize all kinds of the energy densities to the initial photon energy $\rho_{\gamma}(0)=\rho_{\gamma}(N(T_{eq}))$, then we can easily derive the following initial relations,
\begin{alignat}{1}
 &\rho_{\gamma}(0)+\rho_{\nu}(0)=\rho_{B}(0)+\rho_{S}(0),\hspace{0.3cm}
   \rho_{DM}(0)=\rho_{S}(0),\hspace{0.3cm} \rho_{DE}(0)=0\,,\hspace{0.3cm} w_{S}(0)=0\,,\nonumber\\
 &\frac{\rho_{\nu}(0)}{\rho_{\gamma}(0)}=\frac{21}{8}[(\frac{T_{\nu_{L}}}{T_{eq}})^{4}+(\frac{T_{\nu_{R}}}{T_{eq}})^{4}]\,,\hspace{0.3cm}
   \frac{\rho_{B}(0)}{\rho_{\gamma}(0)}=\frac{n_{\gamma}(0)}{\rho_{\gamma}(0)}\frac{n_{B}(0)M_{B}}{n_{\gamma}(0)}
   =\frac{3.6\times10^{10}\eta_{B}\,[\frac{M_{B}}{\mathrm{GeV}}]}{\pi^{4}\,[\frac{T_{eq}}{\mathrm{eV}}]}\,,\nonumber\\
 &\frac{\rho_{S}(0)}{\rho_{\gamma}(0)}=\frac{M_{S}n_{S}(0)}{\rho_{\gamma}(0)}=\frac{\frac{M_{S}}{T_{f}}}{\sqrt{g_{*}(T_{f})}}
   \frac{0.85\times10^{-9}\,\mathrm{GeV}^{-2}}{\langle\sigma v_{r}\rangle_{T_{f}}\,[\frac{T_{eq}}{\mathrm{eV}}]}\,,
\end{alignat}
 where $\frac{T_{\nu_{L}}}{T_{eq}}=(\frac{4}{11})^{\frac{1}{3}}$ and $\frac{T_{\nu_{R}}}{T_{eq}}\approx0.0348^{\frac{1}{3}}$ are given by Eq. (38), $\frac{n_{B}(0)}{n_{\gamma}(0)}=\frac{n_{B}(N_{0})}{n_{\gamma}(N_{0})}=\eta_{B}\approx6.14\times10^{-10}$ is obtained by Eq. (37), and $M_{B}\approx0.9383$ GeV is the baryon mass. A detailed derivation of the $\frac{\rho_{S}(0)}{\rho_{\gamma}(0)}$ formula is seen in Appendix II, on the basis of the discussions of Eq. (8), provided $\sum\limits_{\alpha,\beta}|y^{e}_{\alpha}y^{e}_{\beta}|^{2}\approx1$, $M_{E}\approx0.5$ TeV and $M_{S}\approx256$ GeV, previously we have obtained $\langle\sigma v_{r}\rangle_{T_{f}}\approx1.64\times10^{-9}$ $\mathrm{GeV}^{-2}$, $\frac{M_{S}}{T_{f}}\approx25.4$ and $g_{*}(T_{f})=91.5$. Now put the above values into Eq. (44), thus we can determine $T_{eq}\approx0.928$ eV and $\frac{\rho_{S}(0)}{\rho_{B}(0)}\approx6.46$. Although $\rho_{B}(0)$ and $\rho_{S}(0)$ are the same order of magnitude, we can evidently see from Eq. (44) that they origins are very different. Note that $\frac{\rho_{DM}(N_{0})}{\rho_{B}(N_{0})}\approx5.36<\frac{\rho_{S}(0)}{\rho_{B}(0)}<\frac{\rho_{S}(N_{0})}{\rho_{B}(N_{0})}\approx22.8$, this means that at the present day, only a part of the $S$ particles is still in the CDM phase state, while the supercool part of the $S$ particles has condensed into the dark energy phase state.

 In order to solve Eqs. (40)-(42), we need provide the $\kappa(T)$ evolution. Similar to solving the inflation problem, since the $S$ self-interacting potential is unknown, so I can guess that $\kappa(T)$ is given by the following $F(N)$ function,
\begin{alignat}{1}
 &\kappa(N(T))=\frac{dF(N)}{dN}\;\Longleftrightarrow \int^{N}_{0}\kappa(N')dN'=F(N)=b\,e^{a(1-\frac{N_{0}}{N})},\\
 \Longrightarrow\; &F(0)=0\,,\hspace{0.3cm} F(N_{0})=b\,,\hspace{0.3cm} F(\infty)=b\,e^{a},\nonumber
\end{alignat}
 where $a\approx17.25$ and $b\approx0.186$ are two input parameters in the $S$ condensation process, which are determined by fitting the current density budget of the dark matter and dark energy, see Table 2. The role of Eq. (45) is very similar to that of Eq. (24) discussed in the $\Phi$ inflation, the former controls the rate of $S_{DE}$ growing from $S_{DM}$, whereas the latter characterizes the rate of $\Phi_{DM}$ growing from $\Phi_{DE}$\,.

 Make use of Eqs. (43)- (45), then the relevant energy densities in Eqs. (40)-(42) are analytically solved as follows,
\begin{alignat}{1}
 \frac{dln\rho_{\gamma}}{dN}=-4\Longrightarrow\,&ln\frac{\rho_{\gamma}(N)}{\rho_{\gamma}(0)}=-4N,\nonumber\\
 \frac{dln\rho_{\nu}}{dN}=-3(1+w_{\nu})\Longrightarrow\,&ln\frac{\rho_{\nu}(N)}{\rho_{\gamma}(0)}
                                      =ln\frac{\rho_{\nu}(0)}{\rho_{\gamma}(0)}-4N\:(0\leqslant N\leqslant N_{\nu}),\nonumber\\
 \,&ln\frac{\rho_{\nu}(N)}{\rho_{\gamma}(0)}=ln\frac{\rho_{\nu}(0)}{\rho_{\gamma}(0)}-3N-N_{\nu}\:(N>N_{\nu}),\nonumber\\
 \frac{dln\rho_{B}}{dN}=-3\Longrightarrow\,&ln\frac{\rho_{B}(N)}{\rho_{\gamma}(0)}=ln\frac{\rho_{B}(0)}{\rho_{\gamma}(0)}-3N,\nonumber\\
 \frac{dln\rho_{DM}}{dN}=-3-\kappa\Longrightarrow\,&ln\frac{\rho_{DM}(N)}{\rho_{\gamma}(0)}=ln\frac{\rho_{S}(0)}{\rho_{\gamma}(0)}-3N-F(N),\nonumber\\
 \frac{d\rho_{DE}}{dN}=\kappa\,\rho_{DM}\Longrightarrow\,&ln\frac{\rho_{DE}(N)}{\rho_{\gamma}(0)}
                                   =ln\frac{\rho_{S}(0)}{\rho_{\gamma}(0)}+ln\left[\int_{0}^{N}\kappa(N')e^{-3N'-F(N')}dN'\right],
\end{alignat}
 where $N_{\nu}$ is the time point when the neutrino is transformed from relativistic state to non-relativistic state, see the following Eq. (47), so there is $w_{\nu}=\frac{1}{3}$ when $0\leqslant N\leqslant N_{\nu}$ and $w_{\nu}=0$ when $N>N_{\nu}$\,. Since Eq. (44) has given the initial values, we can immediately calculate from Eq. (46) all kinds of the energy evolutions.

 First of all, we can calculate out the three key time points in the universe evolution,
\ba
 N_{\nu}=ln\frac{T_{eq}}{0.1553\Sigma m_{\nu}}\approx4.52,\hspace{0.3cm} N_{D}=ln\frac{T_{eq}}{3.16\times10^{-4}\,\mathrm{eV}}\approx7.99,\hspace{0.3cm} 
 N_{0}=ln\frac{T_{eq}}{T_{0}}\approx8.28,
\ea
 where $T_{eq}\approx0.928$ eV, its corresponding universe age is about $5\times10^{4}$ years. $N_{\nu}$ is the time at which the neutrino turns into non-relativistic state, its corresponding universe temperature is $0.1553\Sigma m_{\nu}$\,, which is derived from the second equality in Eq. (49), $\Sigma m_{\nu}\approx0.065$ eV is an input parameter of the unified model. $N_{D}$ is the time of the equality of the dark energy density and the total matter one, it occurs at $T\approx3.16\times10^{-4}\,\mathrm{eV}\approx3.67$ K, this is very close to the present-day $T_{0}\approx2.35\times10^{-4}\,\mathrm{eV}\approx2.7255$ K, the present universe age is 13.8 billion years, so the $N_{D}$ time is about 4 billion years ago.

\begin{figure}
 \centering
 \includegraphics[totalheight=7cm]{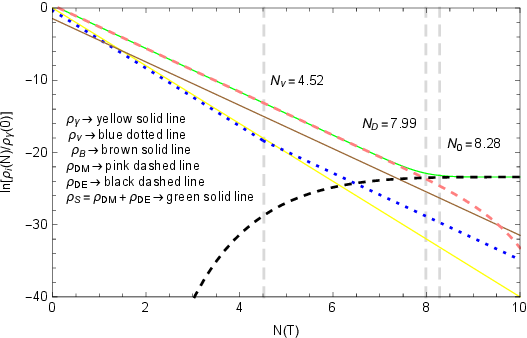}
 \caption{The evolutions of the relevant energy densities with $N(T)$ as time scale since $T_{eq}\approx0.928$ eV. $N(T_{eq})=0$ is set as the starting point of the $S$ condensation, several key time points are also shown. The SM sector performs the normal hot evolution, whereas the dark sector carries out the $S$ condensation, by which $S_{DE}$ is growing from $S_{DM}$, it is essentially a reverse process of the $\Phi$ inflation.}
\end{figure}
 Fig. 8 numerically shows the evolutions of the relevant energy densities since the matter-radiation equality, $N(T_{eq})=0$ is set as the starting point of the $S$ condensation. In the SM sector, $\rho_{\gamma}$ (the yellow curve), $\rho_{\nu}$ (the blue dotted curve) and $\rho_{B}$ (the brown curve) evidently perform the normal hot evolutions. $\rho_{B}$ always keeps the pure matter state since the baryon number is conserved. At $N_{\nu}\approx4.52$, the neutrino turns into the non-relativistic state, accordingly the $\rho_{\nu}$ evolution is transformed from the radiation to the matter, eventually, the radiation is only left with the photon. In the dark sector, $\rho_{S}$ (the green curve), $\rho_{DM}$ (the pink curve) and $\rho_{DE}$ (the black curve) carry out special evolutions. As the universe temperature is more and more cool, more and more DM become supercool so that they condense into the DE, namely, $S_{DE}$ is slowly growing from $S_{DM}$, thereby the total $\rho_{S}$ evolution is gradually deviating from the pure DM and eventually turns into the pure DE. At $N_{D}\approx7.99$, the dark energy exceeds the total matter energy so that it begins to dominate the universe, accordingly the universe is newly transformed from the decelerating expansion to the accelerating one. The present-day value of each energy density is evaluated at $N_{0}\approx8.28$. In the future, once all of the CDM are fully condensed into the dark energy, then the dark energy density will eventually become a constant, thus the universe will become a de Sitter one anew, which is the same state as the primordial universe of the early inflation. When Fig. 8 is compared with Fig. 4, we can see that the current $S$ condensation is essentially a reverse process of the primordial $\Phi$ inflation, the reason for this is of course that $S$ and $\Phi$ are the same type fields and they have the similar nature and dynamics. In a similar way to finding the inflationary potential $V_{\Phi}$, we can also find the condensation potential $V_{S}$, but we will specially discuss it in another paper in order not to increase the length of this paper.

 Further, we can calculate from Eq. (46) the density parameter of each energy component, the total parameter-of-state and the $h$ value, they are given by the following relations,
\begin{alignat}{1}
 &\Omega_{i}(N)=\frac{\rho_{i}(N)}{\sum\limits_{i}\rho_{i}(N)}\,,\hspace{0.3cm}
   w_{T}(N)=\frac{\sum\limits_{i}P_{i}(N)}{\sum\limits_{i}\rho_{i}(N)}=\frac{\Omega_{R}(N)}{3}-\Omega_{DE}(N),\nonumber\\
 &\sum\limits_{i}\rho_{i}(N_{0})=3\tilde{M}^{2}_{p}H_{0}^{2}\;\Longrightarrow h\approx0.73\,,
\end{alignat}
 where the present critical energy density is $3\tilde{M}_{p}^{2}H_{0}^{2}$ with $H_{0}\approx2.13\times10^{-42}h$ GeV, note that $h\approx0.73$ is an output value of the unified model rather than an input parameter. From Eq. (46) and Eq. (48), we can analytically derive the following ratio relations among the present-day density parameters,
\begin{alignat}{1}
 &\Omega_{\gamma}(N_{0})h^{2}=\frac{1}{45}[\frac{\pi\,T^{2}_{0}}{\tilde{M}_{p}\frac{H_{0}}{h}}]^{2}\,,\hspace{0.5cm}
   \frac{\Omega_{\nu}(N_{0})}{\Omega_{\gamma}(N_{0})}=\frac{27\Sigma m_{\nu}}{\pi^{4}T_{0}}
   [(\frac{T_{\nu_{L}}}{T_{0}})^{3}+(\frac{T_{\nu_{R}}}{T_{0}})^{3}],\nonumber\\
 &\frac{\Omega_{B}(N_{0})}{\Omega_{\gamma}(N_{0})}=\frac{36\eta_{B}M_{B}}{\pi^{4}T_{0}}\,,\hspace{0.5cm}
   \frac{\Omega_{DM}(N_{0})}{\Omega_{B}(N_{0})}=\frac{\rho_{S}(0)}{\rho_{B}(0)}\,e^{-F(N_{0})},
\end{alignat}
 where $(\frac{T_{\nu_{L}}}{T_{0}})^{3}=\frac{4}{11}$ and $(\frac{T_{\nu_{R}}}{T_{0}})^{3}=0.0348$. $\Omega_{\gamma}(N_{0})h^{2}\approx2.47\times10^{-5}$ is completely fixed by $T_{0}$, $\Omega_{\nu}(N_{0})$ is determined by $\Sigma m_{\nu}$, and $\Omega_{B}(N_{0})$ are dominated by $\eta_{B}$. Since $\frac{\rho_{S}(0)}{\rho_{B}(0)}\approx6.46$ and $F(N_{0})=b\approx0.186$, then we naturally obtain $\frac{\Omega_{DM}(N_{0})}{\Omega_{B}(N_{0})}\approx5.36$, in addition, $h\approx0.73$ is determined by adjusting the input parameter $a\approx17.25$, finally, $\Omega_{DE}$ is given by the closed relation $\sum\limits_{i}\Omega_{i}\approx1$, the detailed results are all listed in Table 2. 

 Now we make use of my model to solve the ``Hubble tension" problem simply and elegantly, namely we can clarify which one among $h\approx0.73$ and $h\approx0.674$ is really correct. The Planck data analysis of the CMB is on the basis of the following relations \cite{30},
\begin{alignat}{1}
 &r_{s}^{*}=\int_{z_{*}}^{\infty}\frac{c_{s}(z)dz}{H(z)}=\frac{1}{H(z_{*})}\int_{z_{*}}^{\infty}\frac{c_{s}(z)H(z_{*})dz}{H(z)}\,,\hspace{0.3cm}
   H(z_{*})\approx\frac{\sqrt{\rho_{B}(z_{*})[1+\frac{\rho_{DM}(z_{*})}{\rho_{B}(z_{*})}]}}{\sqrt{3}\tilde{M}_{p}}\,,\nonumber\\
 &D_{A}^{*}=\int^{z_{*}}_{0}\frac{dz}{H(z)}=\frac{1}{h\,[\frac{H_{0}}{h}]}\int^{z_{*}}_{0}\frac{H_{0}dz}{H(z)}\,,\hspace{0.5cm}
   \theta^{*}=\frac{r_{s}^{*}}{D_{A}^{*}}\propto\frac{h}{H(z_{*})}\,,
\end{alignat}
 where $z_{*}\approx1100$ is the red-shift from the present day to the photon decoupling, at which the CMB is formed, it is corresponding to the time of $N\approx1.2$ in Fig. 8, namely the universe age is about $3.7\times10^{5}$ years. $r_{s}^{*}$ is the sound horizon at $z_{*}$ and $D_{A}^{*}$ is the angular diameter distance to $z_{*}$. $c_{s}=[3+\frac{\rho_{B}}{\rho_{\gamma}}]^{-\frac{1}{2}}$ is the sound speed of the baryon-photon plasma and it is independent of models, there is generally $\frac{1}{\sqrt{3}}\leqslant c_{s}(z)\lesssim\frac{1}{\sqrt{3.71}}$ for $z\geqslant z_{*}$. $\theta^{*}$ is directly measured from the acoustic peak spacing of the CMB power spectra, so it is a fixed value. In fact, the integral value in the $r_{s}^{*}$ formula is dominated by $\frac{H(z_{*})}{H(z)}\rightarrow1$ when $z\rightarrow z_{*}$, in the same way, the integral value in the $D_{A}^{*}$ formula is dominated by $\frac{H_{0}}{H(z)}\rightarrow1$ when $z\rightarrow 0$, therefore these two factors given by integrating are actually insensitive to the specific models, thus the proportional coefficient of $\theta^{*}\propto\frac{h}{H(z_{*})}$ is almost independent of models. The Planck data analysis assumes the $\Lambda$CDM model, thereby it employs $\frac{\rho_{DM}(z_{*})}{\rho_{B}(z_{*})}=\frac{\rho_{DM}(z)}{\rho_{B}(z)}|_{z=0}\approx5.36$ to infer $h\approx0.674$ by Eq. (50), in contrast, the unified model gives $\frac{\rho_{DM}(z_{*})}{\rho_{B}(z_{*})}\approx\frac{\rho_{DM}(z)}{\rho_{B}(z)}|_{z=z_{eq}}\approx6.46$,  where $z_{eq}\approx3400$ is the red-shift at the R-M equality (namely the time of $N=0$), this result can directly be seen in Fig. 8, thereby we can naturally infer $h\approx0.73$ from Eq. (50). In brief, both of $\theta^{*}\propto\frac{h}{H(z_{*})}\propto\frac{0.674}{\sqrt{1+5.36}}\propto\frac{0.73}{\sqrt{1+6.46}}$ are fitting to the Planck data of the CMB, but the former result is derived from the $\Lambda$CDM model, it is however inconsistent with $h\approx0.73$ measured at the low red-shift, namely there is the ``Hubble tension", whereas the latter result is given by my model, the predicted $h\approx0.73$ is simultaneously fitting to the intersection point given by the S$H_{0}$ES data and the BAO+SNe ones, refer to FIG. 1 in \cite{30}, thus the ``Hubble tension" is naturally eliminated. In conclusion, this unified model solves both the dark energy genesis and the ``Hubble tension", so it is more successful and believable than the $\Lambda$CDM model.

\begin{figure}
 \centering
 \includegraphics[totalheight=7cm]{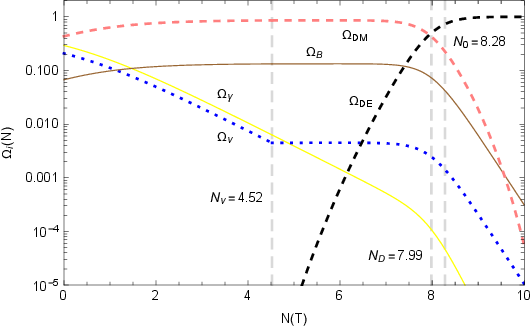}
 \caption{The evolutions of the relevant energy density parameters with $N(T)$ as time scale since $T_{eq}\approx0.928$ eV. $N(T_{eq})=0$ is set as the starting point of the dark energy growing, several key time points are also shown. The present energy density budget is evaluated at $N_{0}$, which is exactly fitting to the present data.}
\end{figure}
 Fig. 9 clearly shows the evolutions of the relevant density parameters since the matter-radiation equality. The standard hot evolution in the visible sector, the CDM condensation and DE genesis in the dark sector are explicitly illustrated by the corresponding curves, these results further confirm our previous discussions. When $N<0$, the universe is R-dominated. When $0<N<N_{D}$, the universe is M-dominated. When $N>N_{D}$, the universe is DE-dominated. The present energy density budget is evaluated at $N_{0}$, which is exactly fitting to the data in Eq. (1), see Table. 2. In the future, the universe will entirely be filled by the dark energy.

\begin{figure}
 \centering
 \includegraphics[totalheight=7cm]{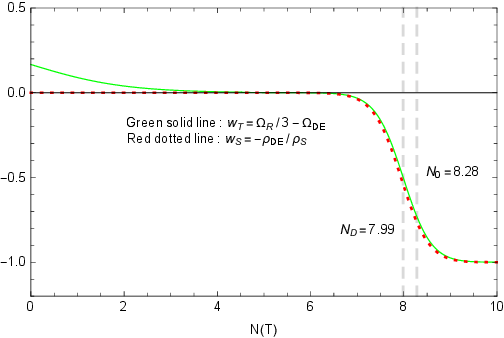}
 \caption{The evolutions of the relevant parameters-of-states with $N(T)$ as time scale since $T_{eq}\approx0.928$ eV. $N(T_{eq})=0$ is set as the starting point of the dark energy growing. At $w_{T}=-\frac{1}{3}$ namely $N\approx7.78$, the universe is newly transformed from the decelerating expansion to the accelerating one. At $w_{T}=-\frac{1}{2}$ namely $N_{D}$, the universe turns into the DE-dominated era. The $S$ condensation is essentially a reverse process of the $\Phi$ inflation. The future fate of the universe will anew become de Sitter universe which is the same state as the primordial universe, but their energy densities differ by about 106 orders of magnitude.}
\end{figure}
 Fig. 10 shows the evolutions of the total parameters-of-state and the $S$ parameters-of-state since the matter-radiation equality. When the $w_{S}$ evolution is compared with the $w_{\Phi}$ evolution given by Fig. 5, we can see that the current CDM condensing into the dark energy is really a reverse process of the primordial slow-roll inflation. When $w_{T}<-\frac{1}{3}$ (namely $N>7.78$), the universe is newly transformed from the decelerating expansion to the accelerating one. When $w_{T}<-\frac{1}{2}$ (namely $N>N_{D}$), the universe is transformed from the M-dominated to the DE-dominated. At the present day (namely $N=N_{0}$), ninety-five percent of the total universe energy is the dark components, and the universe is accelerating expansion by the dark energy driving, all of these have been verified by the observations. The future fate of the universe will anew become de Sitter universe filled with pure $S_{DE}$, which is exactly the same state as the primordial universe filled with pure $\Phi_{DE}$, but these two dark energy densities differ by about 106 orders of magnitude. Up to now, the physical mechanism and picture of the CDM condensation and DE genesis are completely solved. 

\vspace{0.6cm}
\noindent\textbf{VI. Summary for Numerical Results}

\vspace{0.3cm}
 Finally, we summarize all kinds of the important numerical results of the unified model, and then compare them with the measured experimental data. The used physical constants include
\begin{alignat}{1}
 &\tilde{M}_{p}=2.43\times10^{18}\:\mathrm{GeV},\hspace{0.5cm} M_{B}=0.9383\:\mathrm{GeV},\nonumber\\
 &T_{0}=2.7255\:\mathrm{K}=2.35\times10^{-4}\:\mathrm{eV},\hspace{0.3cm}
   \frac{H_{0}}{h}=2.13\times10^{-42}\:\mathrm{GeV}=100\:\mathrm{km}\,\mathrm{s}^{-1}\mathrm{Mpc}^{-1}.
\end{alignat}
 The fundamental quantities of the unified model are set as
\ba
 \mu_{0}\sim M_{N}\sim2\times10^{8}\:\mathrm{GeV},\hspace{0.3cm} v_{\phi}\sim1\:\mathrm{TeV},\hspace{0.3cm} v_{H}\approx246\:\mathrm{GeV},\hspace{0.3cm} v_{\Phi}\sim10\:\mathrm{eV}.
\ea
 In addition, we fix $\sum\limits_{\alpha,\beta}|y^{e}_{\alpha}y^{e}_{\beta}|^{2}\approx1$ and $M_{E}\approx0.5$ TeV in the particle model so that $M_{S}$ becomes an adjustable parameter. Table 2 in detail lists the relevant input parameters and all kinds of the output results in the unified model.
\begin{table}
 \centering
 \begin{tabular}{|c|c|c|c|c|c|c|c|c|}
  \hline\hline
  \multicolumn{9}{|c|}{Relevant input parameters}\\\hline
  \multicolumn{3}{|c|}{$\Phi$ inflation} &\multicolumn{2}{|c|}{$S$ condensation} &\multicolumn{4}{|c|}{Particle model}\\\hline
  $H_{inf}$(GeV) &$\alpha$&$\beta$ &$a$ &$b$ &$\frac{\mu_{0}}{M_{\Phi}}$ &$A_{CP}$ &$M_{S}$(GeV) &$\sum m_{\nu_{i}}$(eV)\\\hline
  $1.99\times10^{10}$ &$3.22$ &$0.3$ &$17.25$ &$0.186$ &$0.005$ &$1.91\times10^{-10}$ &$256$ &$0.065$\\\hline
 \end{tabular}
 \begin{tabular}{|c|c|c|c|c|c|c|c|}
  \hline
  \multicolumn{8}{|c|}{Inflation output quantities}\\\hline
  $k_{*}(\mathrm{Mpc}^{-1})$ &$N_{*}$ &$r_{0.05}$ &$n_{s}$ &$\frac{dn_{s}}{dlnk}$ &$ln[10^{10}\Delta^{2}_{R}]$
  &$\frac{\rho_{\Phi}(N_{*})}{\rho_{\Phi}(t_{inf})}$ &$M_{\Phi}$(GeV)\\\hline
  $0.05$ &$52.43$ &$1.68\times10^{-6}$ &$0.965$ &$-0.0040$ &$3.043$ &$258.9$ &$5.05\times10^{10}$\\\hline
 \end{tabular}
 \begin{tabular}{|c|c|c|c|c|c|}
  \hline
  \multicolumn{6}{|c|}{Reheating output quantities}\\\hline
  &$\Omega_{\Phi}$ &$\Omega_{R}$ &$T$(GeV) &$H$(GeV) &$\Gamma_{\Phi}$(GeV)\\\hline
  $\tilde{t}_{req}=1.054$ &$0.5$ &$0.5$ &$8.29\times10^{10}$ &$1.46\times10^{4}$ &$2.51\times10^{4}$ \\
  $\tilde{t}_{ref}=5$ &$0.019$ &$0.981$ &$4.22\times10^{10}$ &$2.7\times10^{3}$ & \\\hline
 \end{tabular}
 \begin{tabular}{|c|c|c|c|c|c|c|c|}
  \hline
  \multicolumn{8}{|c|}{Current output quantities}\\\hline
  &$\Omega_{\gamma}$ &$\Omega_{\nu}$ &$\Omega_{B}$ &$\Omega_{DM}$ &$\Omega_{DE}$ &$h$ &$\eta_{B}$ \\\hline
  &$4.63\times10^{-5}$ &$1.42\times10^{-3}$ &$0.042$ &$0.2251$ &$0.7315$ &$0.73$ &$6.14\times10^{-10}$ \\
  $\Omega_{i}h^{2}$ &$2.47\times10^{-5}$ &$7.54\times10^{-4}$ &$0.02237$ &$0.120$ &$0.3898$ & & \\\hline\hline
 \end{tabular}
 \caption{A summary of the numerical results of the unified model.}
\end{table}

 In the $\Phi$ inflation there are three input parameters of $H_{inf}$, $\alpha$ and $\beta$, the inflation output results excellently fit all of the inflationary data, moreover, the model predicts $r_{0.05}\approx1.68\times10^{-6}$, which is too small to be detected currently, and the inflaton mass $M_{\Phi}\approx5.05\times10^{10}$ GeV, which purely arises from the inflationary dynamical evolution. The $S$ condensation is controlled by these two input parameters of $a$ and $b$, their values are determined by fitting the current densities of the CDM and dark energy. The remaining input parameters are all provided by the particle model, $\frac{\mu_{0}}{M_{\Phi}}$ and $A_{CP}$ are respectively responsible for the reheating outputs and the baryon asymmetry, $M_{S}$ is in charge of the cross-section and freeze-out temperature of the $S$ annihilation, by which the $S$ relic density is determined, lastly, $\sum m_{\nu_{i}}$ determines the present energy density of the cosmic neutrino. Although there are no observable data of the universe reheating as yet, the reheating output results are all very reasonable and believable. In the last panel, the current energy density budget is perfectly reproduced, moreover, $h\approx0.73$ and $\eta_{B}\approx6.14\times10^{-10}$ is finely predicted, in particular, the ``Hubble tension" is eliminated. In short, all of the numerical results are consistent, reasonable and without any fine-tuning, they are very well in agreement with all of the measured data in Eq. (1).

 All of the energy scales in Table 2 are below the GUT scale, so the primordial inflation really takes place after the GUT phase transition, thus the magnetic monopole problem is naturally eliminated. In the unified model, the slow-roll inflation is driven by the primordial dark energy converting into the super-heavy dark matter, whereas the current dark energy is grown from the CDM condensation, although there is difference about 106 orders of magnitude between these two dark energy densities, the universe energy is step by step released and reduced in the evolution processes of $13.8$ billion years, the whole evolution processes are analogous to a cascade of hydropower stations, see Fig. 3, so there is not the so-called ``cosmological constant energy" at all. All of the above results fully demonstrate that the model is highly of self-consistence, concordance and unification. In conclusion, the unified model is indeed able to account for the universe origin and evolution successfully and excellently, therefore we expect it to be tested by the future experiments.

\vspace{0.6cm}
\noindent\textbf{VII. Conclusions}

\vspace{0.3cm}
 I put forward to a unified model of particle physic and cosmology based on both the new extension of the SM and the fundamental principle of the standard cosmology. This new theory covers the SM physics (visible sector) and the BSM one (dark sector), it can successfully account for the full process of the universe origin and evolution in a unified and integrated way, and also it elegantly explains the origins of the neutrino mass and baryon asymmetry. The primordial universe is a de Sitter one filled with the super-high dark energy $\Phi_{DE}$, the slow-roll inflation is driven by $\Phi_{DE}$ slowly converting into the super-heavy dark matter $\Phi_{DM}$. At the end of the inflation the $\Phi_{DM}$ decay leads to both the reheating and the leptogenesis, after that the universe enters the R-dominated era, and it gradually cools by means of the hot expansion, thus the universe turns into the M-dominated era. The current CDM condensation is driving the $S_{DE}$ dark energy slowly growing from the super-cool dark matter $S_{DM}$. The future universe will anew become a de Sitter one filled with the dark energy $S_{DE}$. In the model, $\Phi$ and $S$ are two special scalar fields with the unusual self-interacting potentials, but they have the similar nature and dynamics, the current $S$ condensation is essentially a reverse process of the primordial $\Phi$ inflation.

 After the new particle model is outlined, for these three process of the inflation, the reheating and the CDM condensation, I give their complete dynamical system of equations and solve their evolutions by some special techniques, in particular, I establish the close relations between these processes and particle physics, which are embodied by some key equations in the paper. The numerical results of the model are in detail shown by all kinds of the figures and Table 2, these results clearly illustrate the physical mechanism and picture of each process. For the slow-roll inflation, I in detail discuss the conversion among the energy forms, the inflaton mass generation, the inflationary potential profile and its function form. In addition, I deeply study the details of the reheating evolution, baryogenesis, CDM condensation and dark energy genesis, and predict the future fate of the universe. From the super-high primordial dark energy to the super-low current dark energy, the universe energy is step by step released and reduced in the evolution history of $13.8$ billion years, which is analogous to a cascade of hydropower stations, so there is no the ``cosmological constant energy".

 The unified model can excellently fit all kinds of the observed data by use of fewer input parameters. It not only perfectly reproduces the measured inflationary data and the current energy density budget, but also finely predicts some important cosmological quantities such as the tensor-to-scalar ratio $r_{0.05}\approx1.68\times10^{-6}$, the inflaton mass $M_{\Phi}\approx5.05\times10^{10}$ GeV, the reheating temperature $T_{re}\approx8.29\times10^{10}$ GeV, the CDM mass $M_{S}\approx256$ GeV, the baryon asymmetry $\eta_{B}\approx6.14\times10^{-10}$, the scaling factor of expansion rate $h\approx0.73$, in particular, it also clarifies and eliminates the ``Hubble tension". In short, the model really achieves a unification of particle physics and cosmology, all of the given results are very significant and believable, therefore we expect to test the unified model in the near future.

\vspace{0.6cm}
 \noindent\textbf{Acknowledgements}

\vspace{0.3cm}
 I would like to thank my wife for her great helps. This research is supported by the Fundamental Research Funds for the Central Universities of China under Grant No. WY2030040065.

\vspace{0.6cm}
\noindent\textbf{Appendix I}

\vspace{0.3cm}
 A derivation of the $k_{*}$ formula in Eq. (28) is as follows,
\begin{alignat}{1}
 k_{*}&=\frac{a_{*}H_{*}}{c}=\frac{H_{0}}{c}\frac{H_{inf}}{H_{0}}\frac{H_{*}}{H_{inf}}
              \frac{a_{*}}{a_{inf}}\frac{a_{inf}}{a_{req}}\frac{a_{req}}{a_{ref}}\frac{a_{ref}}{a_{0}}\nonumber\\
 &=\frac{H_{0}}{c}\frac{H_{inf}}{H_{0}}[\frac{\rho_{\Phi}(N_{*})}{\rho_{\Phi}(0)}]^{\frac{1}{2}}e^{-N_{*}}
     [\frac{\rho_{\Phi}(t_{req})}{\rho_{\Phi}(t_{inf})}]^{\frac{1}{3}}
     [\frac{\rho_{R}(t_{ref})}{\rho_{R}(t_{req})}]^{\frac{1}{4}}[\frac{s(T_{0})}{s(T_{ref})}]^{\frac{1}{3}}\nonumber\\
 &=\frac{H_{0}}{c}\frac{H_{inf}}{H_{0}}[\frac{\rho_{\Phi}(N_{*})}{\rho_{\Phi}(0)}]^{\frac{1}{2}}e^{-N_{*}}
     [\frac{\rho_{c}(t_{0})}{\rho_{\Phi}(t_{inf})}\frac{\rho_{\gamma}(t_{0})}{\rho_{c}(t_{0})}\frac{\rho_{R}(t_{req})}{\rho_{\gamma}(t_{0})}]^{\frac{1}{3}}
     [\frac{T_{ref}}{T_{req}}][\frac{g_{*}^{\frac{1}{3}}(T_{0})T_{0}}{g_{*}^{\frac{1}{3}}(T_{ref})T_{ref}}]\nonumber\\
 &=\frac{H_{0}}{c}\frac{H_{inf}}{H_{0}}[\frac{\rho_{\Phi}(N_{*})}{\rho_{\Phi}(0)}]^{\frac{1}{2}}e^{-N_{*}}
     [\frac{H_{0}^{2}}{H_{inf}^{2}}]^{\frac{1}{3}}[\Omega_{\gamma}(T_{0})]^{\frac{1}{3}}
     [\frac{g_{*}(T_{req})T^{4}_{req}}{2T_{0}^{4}}]^{\frac{1}{3}}[\frac{g_{*}^{\frac{1}{3}}(T_{0})T_{0}}{g_{*}^{\frac{1}{3}}(T_{ref})T_{req}}]\nonumber\\
 &=[\frac{H_{0}}{c\,h}][\frac{g_{*}(T_{0})}{2}]^{\frac{1}{3}}[\Omega_{\gamma}(T_{0})h^{2}]^{\frac{1}{3}}
     [\frac{H_{inf}}{H_{0}/h}]^{\frac{1}{3}}[\frac{T_{req}}{T_{0}}]^{\frac{1}{3}}[\frac{\rho_{\Phi}(N_{*})}{\rho_{\Phi}(0)}]^{\frac{1}{2}}e^{-N_{*}},
\end{alignat}
 where I use $a\propto\rho^{-\frac{1}{3}}$ for the $\Phi$-dominated phase in $t_{inf}<t<t_{req}$ and $a\propto\rho^{-\frac{1}{4}}$ for the R-dominated phase in $t_{req}<t<t_{ref}$, in addition, I employ $\rho_{\Phi}(t_{req})=\rho_{R}(t_{req})$ and $g_{*}(T_{req})=g_{*}(T_{ref})$.

\vspace{0.6cm}
\noindent\textbf{Appendix II}

\vspace{0.3cm}
 A derivation of the $\frac{\rho_{S}(0)}{\rho_{\gamma}(0)}$ formula in Eq. (44) is as follows,
\begin{alignat}{1}
 &\frac{\rho_{S}(0)}{\rho_{\gamma}(0)}=\frac{M_{S}n_{S}(0)}{\rho_{\gamma}(0)}
   =\frac{M_{S}n_{S}(T_{f})}{\rho_{\gamma}(T_{eq})}\frac{n_{S}(T_{eq})}{n_{S}(T_{f})}\,,\hspace{0.3cm}
   \frac{n_{S}(T_{eq})}{n_{S}(T_{f})}=\frac{a^{3}(T_{f})}{a^{3}(T_{eq})}=\frac{g_{*}(T_{eq})T_{eq}^{3}}{g_{*}(T_{f})T_{f}^{3}}\,,\nonumber\\
 &n_{S}(T_{f})=\frac{H(T_{f})}{\langle\sigma v_{r}\rangle_{T_{f}}}
   =\frac{1.66\sqrt{g_{*}(T_{f})}\,T_{f}^{2}}{\langle\sigma v_{r}\rangle_{T_{f}}M_{Pl}}\,,\hspace{0.3cm}
   \rho_{\gamma}(T_{eq})=\frac{\pi^{2}}{15}\,T_{eq}^{4}\,,\nonumber\\
 &\Longrightarrow \frac{\rho_{S}(0)}{\rho_{\gamma}(0)}=\frac{M_{S}}{T_{f}\sqrt{g_{*}(T_{f})}}
   \frac{8.5\times10^{-10}\,\mathrm{GeV}^{-2}}{\langle\sigma v_{r}\rangle_{T_{f}}[\frac{T_{eq}}{\mathrm{eV}}]}\,,
\end{alignat}
 where $g_{*}(T_{eq})=4.1$ and $g_{*}(T_{f})=91.5$\,.

\end{document}